\documentclass[twocolumn]{pasj01}

\RequirePackage{mathpazo}
\RequirePackage[T1]{fontenc}

\usepackage[switch,mathlines]{lineno}
\usepackage{comment}
\usepackage{lscape}
\usepackage{xcolor}

\def\commenta{$^*$}
\def\commentb{$^\dagger$}
\def\commentc{$^\ddagger$}

\newcounter{author}
\def\authorcount#1#2{\refstepcounter{author}\label{#1}
\altaffiltext{\ref{#1}}{#2}}

\Received{$\langle$reception date$\rangle$}
\Accepted{$\langle$acception date$\rangle$}
\Published{$\langle$publication date$\rangle$}

\begin{document}
\SetRunningHead{Y. Tampo et. al.}{MASTER OT J030227.28+191754.5 2021 superoutburst}

\title{MASTER OT J030227.28+191754.5: an unprecedentedly energetic dwarf nova outburst}

\author{
    Yusuke~\textsc{Tampo}\altaffilmark{\ref{affil:Kyoto}*}$^,$\altaffilmark{\ref{affil:saao}}$^,$\altaffilmark{\ref{affil:uct}},
    Taichi~\textsc{Kato}\altaffilmark{\ref{affil:Kyoto}},
    Keisuke~\textsc{Isogai}\altaffilmark{\ref{affil:KyotoOkayama}}$^,$\altaffilmark{\ref{affil:utokyokomaba}},
    Mariko~\textsc{Kimura}\altaffilmark{\ref{affil:riken}},
    Naoto~\textsc{Kojiguchi}\altaffilmark{\ref{affil:Kyoto}},
    Daisaku~\textsc{Nogami}\altaffilmark{\ref{affil:Kyoto}},
    Junpei~\textsc{Ito}\altaffilmark{\ref{affil:Kyoto}},
    Masaaki~\textsc{Shibata}\altaffilmark{\ref{affil:Kyoto}},
    Masayuki~\textsc{Yamanaka}\altaffilmark{\ref{affil:kagoshima}}$^,$\altaffilmark{\ref{affil:KyotoOkayama}},
    Kenta~\textsc{Taguchi}\altaffilmark{\ref{affil:Kyoto}},
    Hiroyuki~\textsc{Maehara}\altaffilmark{\ref{affil:NAOJOkayama}},
    Hiroshi~\textsc{Itoh}\altaffilmark{\ref{affil:Ioh}}, 
    Katsura~\textsc{Matsumoto}\altaffilmark{\ref{affil:oku}}, 
    Momoka~\textsc{Nakagawa}\altaffilmark{\ref{affil:oku}}, 
    Yukitaka~\textsc{Nishida}\altaffilmark{\ref{affil:oku}}, 
    Shawn~\textsc{Dvorak}\altaffilmark{\ref{affil:dks}}, 
    Katsuhiro~L.~\textsc{Murata}\altaffilmark{\ref{affil:mtm}}, 
    Ryohei~\textsc{Hosokawa}\altaffilmark{\ref{affil:mtm}}, 
    Yuri~\textsc{Imai}\altaffilmark{\ref{affil:mtm}}, 
    Naohiro~\textsc{Ito}\altaffilmark{\ref{affil:mtm}}, 
    Masafumi~\textsc{Niwano}\altaffilmark{\ref{affil:mtm}}, 
    Shota~\textsc{Sato}\altaffilmark{\ref{affil:mtm}}, 
    Ryotaro~\textsc{Noto}\altaffilmark{\ref{affil:mtm}}, 
    Ryodai~\textsc{Yamaguchi}\altaffilmark{\ref{affil:mtm}}, 
    Malte~\textsc{Schramm}\altaffilmark{\ref{affil:NAOJ}}$^,$\altaffilmark{\ref{affil:saitamau}},
    Yumiko~\textsc{Oasa}\altaffilmark{\ref{affil:saitamau}}, 
    Takahiro~\textsc{Kanai}\altaffilmark{\ref{affil:saitamau}}, 
    Yu~\textsc{Sasaki}\altaffilmark{\ref{affil:saitamau}}, 
    Tam\'as~\textsc{Tordai}\altaffilmark{\ref{affil:trt}},  
    Tonny~\textsc{Vanmunster}\altaffilmark{\ref{affil:Van1}}$^,$\altaffilmark{\ref{affil:Van2}}, 
    Seiichiro~\textsc{Kiyota}\altaffilmark{\ref{affil:kis}},   
    Nataly~\textsc{Katysheva}\altaffilmark{\ref{affil:saimsu}},
    Sergey~Yu.~\textsc{Shugarov}\altaffilmark{\ref{affil:shu2}}$^,$\altaffilmark{\ref{affil:saimsu}},  
    Alexandra~M.~\textsc{Zubareva}\altaffilmark{\ref{affil:saimsu}}$^,$\altaffilmark{\ref{affil:zub}}, 
    Sergei~\textsc{Antipin}\altaffilmark{\ref{affil:saimsu}}, 
    Natalia~\textsc{Ikonnikova}\altaffilmark{\ref{affil:saimsu}}, 
    Alexandr~\textsc{Belinski}\altaffilmark{\ref{affil:saimsu}}, 
    Pavol~A.~\textsc{Dubovsky}\altaffilmark{\ref{affil:Vih}}, 
    Tom\'a\v{s}~\textsc{Medulka}\altaffilmark{\ref{affil:Vih}}, 
    Jun~\textsc{Takahashi}\altaffilmark{\ref{affil:nhao}}, 
    Masaki~\textsc{Takayama}\altaffilmark{\ref{affil:nhao}},
    Tomohito~\textsc{Ohshima}\altaffilmark{\ref{affil:nhao}},
    Tomoki~\textsc{Saito}\altaffilmark{\ref{affil:nhao}}, 
    Miyako~\textsc{Tozuka}\altaffilmark{\ref{affil:nhao}},
    Shigeyuki~\textsc{Sako}\altaffilmark{\ref{affil:utokyomitaka}}$^,$\altaffilmark{\ref{affil:utokyoplanetary}}$^,$\altaffilmark{\ref{affil:sako1}},
    Masaomi~\textsc{Tanaka}\altaffilmark{\ref{affil:tohokuu}}$^,$\altaffilmark{\ref{affil:tohokuu1}},
    Nozomu~\textsc{Tominaga}\altaffilmark{\ref{affil:NAOJ}}$^,$\altaffilmark{\ref{affil:sokendai}}$^,$\altaffilmark{\ref{affil:konanu}}$^,$\altaffilmark{\ref{affil:kavli}},  
    Takashi~\textsc{Horiuchi}\altaffilmark{\ref{affil:utokyomitaka}}$^,$\altaffilmark{\ref{affil:ish}},
    Hidekazu~\textsc{Hanayama}\altaffilmark{\ref{affil:ish}}, 
    Daniel~E.~\textsc{Reichart}\altaffilmark{\ref{affil:skynet}}, 
    Vladimir~V.~\textsc{Kouprianov}\altaffilmark{\ref{affil:skynet}},
    James~W.~\textsc{Davidson}~Jr\altaffilmark{\ref{affil:rrrt}}, 
    Daniel~B.~\textsc{Caton}\altaffilmark{\ref{affil:dso17}}, 
    Filipp~D.~\textsc{Romanov}\altaffilmark{\ref{affil:RFD}}$^,$\altaffilmark{\ref{affil:ARO}}$^,$\altaffilmark{\ref{affil:AAVSO}}, 
    David~J.~\textsc{Lane}\altaffilmark{\ref{affil:ARO}}$^,$\altaffilmark{\ref{affil:AAVSO}}, 
    Franz-Josef~\textsc{Hambsch}\altaffilmark{\ref{affil:ham1}}$^,$\altaffilmark{\ref{affil:ham2}}$^,$\altaffilmark{\ref{affil:dfs}}, 
    Norio~\textsc{Narita}\altaffilmark{\ref{affil:Komaba2}}$^,$\altaffilmark{\ref{affil:ABC}}$^,$\altaffilmark{\ref{affil:IAC_spain}},
    Akihiko~\textsc{Fukui}\altaffilmark{\ref{affil:Komaba2}}$^,$\altaffilmark{\ref{affil:IAC_spain}},
    Masahiro~\textsc{Ikoma}\altaffilmark{\ref{affil:NAOJscience}}, 
    Motohide~\textsc{Tamura}\altaffilmark{\ref{affil:UTokyo}}$^,$\altaffilmark{\ref{affil:ABC}}$^,$\altaffilmark{\ref{affil:NAOJ}},
    Koji~S.~\textsc{Kawabata}\altaffilmark{\ref{affil:hho}}, 
    Tatsuya~\textsc{Nakaoka}\altaffilmark{\ref{affil:hho}}, and 
    Ryo~\textsc{Imazawa}\altaffilmark{\ref{affil:hiroshimaU}} 
}

\authorcount{affil:Kyoto}{
     Department of Astronomy, Kyoto University, Kitashirakawa-Oiwake-cho, Sakyo-ku, 
     Kyoto 606-8502, Japan}
\email{$^*$tampo@kusastro.kyoto-u.ac.jp}

\authorcount{affil:saao}{
    South African Astronomical Observatory, PO Box 9, Observatory, 7935, Cape Town, South Africa}
\authorcount{affil:uct}{
    Department of Astronomy, University of Cape Town, Private Bag X3, Rondebosch 7701, South Africa}

\authorcount{affil:KyotoOkayama}{
     Okayama Observatory, Kyoto University, 3037-5 Honjo, Kamogatacho,
     Asakuchi, Okayama 719-0232, Japan}

\authorcount{affil:utokyokomaba}{
Department of Multi-Disciplinary Sciences, Graduate School of Arts and Sciences, The
University of Tokyo, 3-8-1 Komaba, Meguro, Tokyo 153-8902, Japan}

\authorcount{affil:riken}{
     Cluster for Pioneering Research, Institute of Physical and Chemical Research (RIKEN), 2-1
    Hirosawa, Wako, Saitama 351-0198}

\authorcount{affil:kagoshima}{
Amanogawa Galaxy Astronomy Research Center, Graduate School of Science and Engineering, Kagoshima University,
1-21-35 Korimoto, Kagoshima 890-0065, Japan}

\authorcount{affil:NAOJOkayama}{
     Subaru Telescope Okayama Branch Office, National Astronomical Observatory of Japan, 
     National Institutes of Natural Sciences, 3037-5 Honjo, Kamogata, Asakuchi, Okayama 719-0232, Japan}

\authorcount{affil:Ioh}{
     Variable Star Observers League in Japan (VSOLJ),
     1001-105 Nishiterakata, Hachioji, Tokyo 192-0153, Japan}

\authorcount{affil:oku}{
Osaka Kyoiku University, 4-698-1 Asahigaoka, Osaka 582-8582, Japan}

\authorcount{affil:dks}{
Rolling Hills Observatory, 1643 Nightfall Drive, Clermont, Florida 34711, USA}

\authorcount{affil:mtm}{
Department of Physics, Tokyo Institute of Technology, 2-12-1 Ookayama, Meguro-ku, Tokyo
152-8551, Japan}

\authorcount{affil:NAOJ}{
    National Astronomical Observatory of Japan, National Institutes of Natural Sciences,
    2-21-1 Osawa, Mitaka, Tokyo 181-8588, Japan}

\authorcount{affil:saitamau}{
Graduate school of Science and Engineering, Saitama Univ. 255 Shimo-Okubo, Sakura-ku, Saitama City, Saitama 338-8570, Japan}

\authorcount{affil:trt}{
     Polaris Observatory, Hungarian Astronomical Association,
     Laborc utca 2/c, 1037 Budapest, Hungary}

\authorcount{affil:Van1}{
     Center for Backyard Astrophysics Belgium, 
     Walhostraat 1a, B-3401 Landen, Belgium}

\authorcount{affil:Van2}{
    Center for Backyard Astrophysics Extremadura, 
    e-EyE Astronomical Complex, 
    ES-06340 Fregenal de la Sierra, Spain}

\authorcount{affil:kis}{
Variable Star Observers League in Japan (VSOLJ), 7-1 Kitahatsutomi, Kamagaya, Chiba 273-0126, Japan}

\authorcount{affil:saimsu}{
    Sternberg Astronomical Institute, Lomonosov Moscow State University, Universitetsky Ave.,13, Moscow 119234, Russia}

\authorcount{affil:shu2}{
Astronomical Institute of the Slovak Academy of Sciences, 05960 Tatransk\'a Lomnica, Slovakia}

\authorcount{affil:zub}{
    Institute of Astronomy, Russian Academy of Sciences, Moscow 119017, Russia}    

\authorcount{affil:Vih}{
     Vihorlat Observatory, Mierova 4, 06601 Humenne, Slovakia}

\authorcount{affil:nhao}{
Center for Astronomy, University of Hyogo, 407-2 Nishigaichi, Sayo, Hyogo 679-5313, Japan}

\authorcount{affil:utokyomitaka}{
Institute of Astronomy, Graduate School of Science, The University of Tokyo, 2-21-1 Osawa, Mitaka,
Tokyo 181-0015, Japan}

\authorcount{affil:utokyoplanetary}{
UTokyo Organization for Planetary Space Science, The University of Tokyo, 7-3-1 Hongo, Bunkyo-ku,
Tokyo 113-0033, Japan}

\authorcount{affil:sako1}{
Collaborative Research Organization for Space Science and Technology, The University of Tokyo, 7-3-1
Hongo, Bunkyo-ku, Tokyo 113-0033, Japan}

\authorcount{affil:tohokuu}{Astronomical Institute, Tohoku University, Aoba, Sendai 980-8578, Japan}

\authorcount{affil:tohokuu1}{Division for the Establishment of Frontier Sciences, Organization for Advanced Studies, Tohoku University, Sendai 980-8577, Japan}

\authorcount{affil:sokendai}{
Department of Astronomical Science, School of Physical Sciences, The Graduate University of Advanced Studies (SOKENDAI), 2-21-1 Osawa, Mitaka, Tokyo 181-8588, Japan}

\authorcount{affil:konanu}{
Department of Physics, Faculty of Science and Engineering, Konan University, 8-9-1 Okamoto, Kobe, Hyogo 658-8501, Japan}

\authorcount{affil:kavli}{
Kavli Institute for the Physics and Mathematics of the Universe (WPI), The University of Tokyo, 5-1-5 Kashiwanoha, Kashiwa, Chiba 277-8583, Japan }

\authorcount{affil:ish}{
Ishigakijima Astronomical Observatory,
Public Relations Center,
National Astronomical Observatory of Japan,
1024-1 Arakawa, Ishigaki, Okinawa, 907-0024, Japan}

\authorcount{affil:skynet}{
Department of Physics and Astronomy, University of North Carolina at Chapel Hill, Campus Box 3255, Chapel Hill, NC 27599-3255, USA}

\authorcount{affil:rrrt}{
Department of Astronomy, University of Virginia, P.O. Box 400325, Charlottesville, VA 22904, USA}

\authorcount{affil:dso17}{Department of Physics and Astronomy, Appalachian State University, Boone, NC 28608, USA}

\authorcount{affil:RFD}{
Pobedy street, house 7, flat 60, Yuzhno-Morskoy, Nakhodka, Primorsky
Krai 692954, Russia}

\authorcount{affil:ARO}{
Abbey Ridge Observatory, 45 Abbey Rd, Stillwater Lake, NS, B3Z1R1 Canada}

\authorcount{affil:AAVSO}{
American Association of Variable Star Observers (AAVSO), 185 Alewife
Brook Parkway, Suite 410, Cambridge, MA 02138, USA}

\authorcount{affil:ham1}{
    Groupe Européen d’Observations Stellaires (GEOS), 23 Parc de Levesville, 28300 Bailleau l’Evêque, France}
\authorcount{affil:ham2}{
    Bundesdeutsche Arbeitsgemeinschaft für Veränderliche Sterne (BAV), Munsterdamm 90, 12169 Berlin, Germany}

\authorcount{affil:dfs}{
    Vereniging Voor Sterrenkunde (VVS), Oostmeers 122 C, 8000 Brugge, Belgium}

\authorcount{affil:Komaba2}{
Komaba Institute for Science, The University of Tokyo, 3-8-1 Komaba, Meguro, Tokyo 153-8902, Japan
}

\authorcount{affil:ABC}{
    Astrobiology Center, National Institutes of Natural Sciences, 
    2-21-1 Osawa, Mitaka, Tokyo 181-8588, Japan}

\authorcount{affil:IAC_spain}{
    Instituto de Astrof\'isica de Canarias, V\'ia L\'actea s/n,
    E-38205 La Laguna, Tenerife, Spain}

\authorcount{affil:NAOJscience}{
Division of Science, National Astronomical Observatory of Japan, 2-21-1 Osawa, Mitaka, Tokyo 181-8588, Japan
}

\authorcount{affil:UTokyo}{
    Department of Astronomy, Graduate School of Science, 
    The University of Tokyo, 7-3-1 Hongo, Bunkyo-ku, Tokyo 113-0033, Japan}

\authorcount{affil:hho}{
Hiroshima Astrophysical Science Center, Hiroshima University, Kagamiyama 1-3-1, Higashi-Hiroshima, Hiroshima 739-8526, Japan
}

\authorcount{affil:hiroshimaU}{
Department of Physics, Graduate School of Advanced Science and Engineering, Hiroshima University Kagamiyama, 1-3-1 Higashi-Hiroshima, Hiroshima 739-8526, Japan
}



\maketitle

\begin{abstract}

We present a detailed study of the MASTER OT J030227.28+191754.5 outburst in 2021-2022, 
reaching an amplitude of 10.2 mag and a duration of 60 d. The detections of (1) the double-peaked optical emission lines, and (2) the early and ordinary superhumps, established that  MASTER OT J030227.28+191754.5 is an extremely energetic WZ Sge-type dwarf nova (DN). Based on the superhump observations, we obtained its orbital period and mass ratio as 0.05986(1) d and 0.063(1), respectively. These are within a typical range of low-mass-ratio DNe. 
According to the binary parameters derived based on the thermal-tidal instability model, our analyses showed that (1) the standard disk model requires an accretion rate $\simeq$ 10$^{20}$ g s$^{-1}$ to explain its peak optical luminosity and (2) large mass was stored in the disk at the outburst onset. These cannot be explained solely by the impact of its massive ($\gtrsim$ 1.15 M$_\odot$) primary white dwarf implied by \citet{kim23j0302}.
Instead, we propose that the probable origin of this enormously energetic DN outburst is the even lower quiescence viscosity than other WZ Sge-type DNe. This discussion is qualitatively valid for most possible binary parameter spaces unless the inclination is low ($\lesssim 40^\circ$) enough for the disk to be bright explaining the outburst amplitude. Such low inclinations, however, would not allow detectable amplitude of early superhumps in the current thermal-tidal instability model.
The optical spectra at outburst maximum showed the strong emission lines of Balmer, He~\textsc{i}, and He~\textsc{ii} series whose core is narrower than $\sim 800$ km s$^{-1}$.
Considering its binary parameters, a Keplerian disk cannot explain this narrow component, but the presumable origin is disk winds.

\end{abstract}


\section{Introduction}
\label{sec:1}

Cataclysmic variables (CVs) are close binary systems composed of a primary white dwarf (WD), a low-mass secondary star that fills its Roche lobe, and an accretion disk around the WD [see \citet{war95book,hel01book} for a general review]. Dwarf novae (DNe), a subclass of CVs, show recurrent outbursts in an accretion disk. The mechanism of DN outbursts is understood in the thermal instability model in a disk \citep{osa96review, kim20thesis}, in which viscosity jump between neutral and ionized hydrogen gas triggers an increase of the disk accretion rate, and the released gravitational energy is observed as an outburst.

WZ Sge-type DNe forms a subclass in DNe, whose mass ratios are typically below 0.1 (see \cite{kat15wzsge}). These show most energetic outbursts in DNe; a duration of a few weeks, an outburst amplitude up to nine magnitudes, and an outburst cycle reaching a decade or even longer. The distinctive feature of WZ Sge-type DNe in outburst is early and ordinary superhumps. Early superhumps show a double-wave profile with a stable period almost identical to its orbital period ($\sim$0.1\% accuracy; \cite{pat81wzsge,ish02wzsgeletter}), whereas ordinary superhumps are a single-peaked variation with periods a few percent longer than its orbital one (e.g., \cite{Pdot}). The outbursts accompanying superhumps are called superoutbursts, and these are explained in the thermal--tidal instability (TTI) model \citep{osa89suuma}. Provided a low mass ratio of WZ Sge-type DNe, the disk reaching the 2:1 and 3:1 resonance radius deforms into a vertically-extended double spiral pattern \citep{lin79lowqdisk, kun05earySHSPH} and eccentric shape \citep{whi88ADsimulation2, hir90SHexcess}, which is observed as early and ordinary superhumps, respectively. Early superhumps are always observed before ordinary superhumps \citep{lub91SHa}. The ordinary superhump phase is systematically divided into three stages which is interpreted as the propagation of the eccentricity in a disk; stage A superhump with the longest and constant period, stage B superhump with a positive $P_{\rm dot}$ ($=\dot{P}/P$), and stage C superhump with a shorter superhump period \citep{Pdot, kat13qfromstageA}. Furthermore, in the TTI model, energetic superoutbursts in WZ Sge-type DNe are understood as a result of a relatively lower mass-transfer rate and lower quiescence viscosity than other subtypes of DNe \citep{sma93wzsge, osa95wzsge}.  We note that there is another proposed mechanism to explain DN outbursts in which the mass transfer burst from the secondary star triggers an outburst (enhanced mass transfer model; e.g., \cite{ham20CVreview}).

In this paper, we present our optical -- near-infrared (IR) observation campaign of MASTER OT J030227.28+191754.5 during its 2021-2022 superoutburst. This outburst reached an amplitude of 10.2 mag and a duration of 60 d, enormously energetic as a DN outburst. We examine how the widely-accepted TTI model can explain these outburst properties. 
The rest of this paper is organized as follows:
We introduce the basic information of MASTER OT J030227.28+191754.5 in section \ref{sec:2}. Section \ref{sec:3} presents the overview of our photometric and spectroscopic observations in optical and IR wavelength. Our results from these observations are presented in sections \ref{sec:4} and \ref{sec:5}. In section \ref{sec:6}, we derive the binary parameters of MASTER J0302. We discuss the possible nature of MASTER OT J030227.28+191754.5 in section \ref{sec:7} and summarize this paper in section \ref{sec:8}.

\section{MASTER OT J030227.28+191754.5}
\label{sec:2}

MASTER OT J030227.28+191754.5 (= AT 2021afpi\footnote{https://www.wis-tns.org/object/2021afpi} = PNV J03022732$+$1917552\footnote{http://www.cbat.eps.harvard.edu/unconf/followups/J03022732+1917552.html}, hereafter MASTER J0302) was initially discovered by \citet{zhi21J0302discovery} on 2021-11-26.82568 UT (= BJD 2459545.331) during a search for a corresponding optical transient of the neutrino event Icecube-211125A found on BJD 2459543.771034 \citep{Ice21J0302Icecubegcn31126}.
The follow-up observations have detected double-peaked emission lines in the optical spectrum \citep{iso21J0302specatel15074}, early superhumps (vsnet-alert 26477\footnote{http://ooruri.kusastro.kyoto-u.ac.jp/mailarchive/vsnet-alert/26477}), and ordinary superhumps (vsnet-alert 26501\footnote{http://ooruri.kusastro.kyoto-u.ac.jp/mailarchive/vsnet-alert/26501}), confirming MASTER J0302 as a WZ Sge-type DN outburst.
\citet{sar21J0302predicatel15081} reported that MASTER J0302 on BJD 2459543.4, before Icecube-211125A, was already 1-mag brighter than its quiescence. Based on this fact, they suggested that MASTER J0302 is presumably not associated with the neutrino event.
We note that, in passing, \citet{pai21J0302atel15085} later reported a possible association of Icecube-211125A with a BL Lacertae object 4FGLJ0258.1+2030.

The apparent magnitudes of its quiescence counterpart are summarized in table \ref{tab:quiescencemag}. There is a nearby star (Gaia DR3 59985471761207168) with a separation of 1.7 arcsec from MASTER J0302, and the contaminated data are visually excluded. Although there is no available parallax of MASTER J0302 in Gaia DR3 \citep{gaiadr3}, \citet{kat22WZSgecandle} estimated its distance as 720 pc by applying the fact that the appearance of ordinary superhumps can be used as a standard candle. The Galactic reddening in the direction of MASTER J0302 is estimated as $E(g-r) = 0.13(2)$  at a distance 300--4000 pc \citep{gre19dustextinction}. The extinction of each band was calculated based on \citet{fit99extinction,sch11extinction,gre19dustextinction,wal19galexectionction} assuming standard  $R_{V} = 3.1$. The de-reddened quiescence magnitudes are also listed in table \ref{tab:quiescencemag}.

\begin{table}
\caption{The apparent and de-reddened magnitude of quiescence counterpart of MASTER J0302}
\centering
\label{tab:quiescencemag}
\begin{tabular}{ccc}
  \hline              
    band & apparent (mag) & de-reddened (mag) \\
    \hline
    $FUV$\commenta   & $21.35\pm0.29$ & $20.19\pm0.34$ \\ 
    $NUV$\commenta   & $21.50\pm0.28$ & $20.52\pm0.31$ \\ 
    $u$\commentb     & $22.17\pm0.19$ & $21.55\pm0.21$ \\ 
    $g$\commentb     & $21.94\pm0.07$ & $21.46\pm0.10$ \\ 
    $r$\commentb     & $21.93\pm0.10$ & $21.59\pm0.11$ \\ 
    $i$\commentb     & $22.19\pm0.19$ & $21.95\pm0.20$ \\ 
    $z$\commentb     & $22.42\pm0.58$ & $22.23\pm0.58$ \\ 
  \hline
    \multicolumn{3}{l}{\commenta GALEX J030227.3+191754, }\\
    \multicolumn{3}{l}{data from GALEX-DR5 \citep{galexdr5}.}\\
    \multicolumn{3}{l}{\commentb SDSS J030227.29+191754.7,}\\
    \multicolumn{3}{l}{data from SDSS DR16 \citep{sdssdr16}.}\\
\end{tabular}
\end{table}

In addition to the optical-IR data, MASTER J0302 was observed in X-rays with {\em NICER} and {\em NuSTAR}  during the superoutburst. These results are presented in \citet{kim23j0302}, and we briefly summarize their important results. The X-ray spectrum on BJD 2459551 (i.e., around outburst maximum) showed strong emission lines of oxygen and neon. Their yielded abundance was several times higher than the solar values, indicating that MASTER J0302 hosts an Oxygen-Neon (ONe) WD. Moreover, the X-ray spectrum showed blackbody emission at a temperature of $\sim 30$ eV. Assuming the binary parameters estimated from the TTI model and interpreting this as emission from the belt-shaped optically-thick boundary layer,  \citet{kim23j0302} estimated the WD radius as $(2.9\pm{1.1})\times10^{8}$ cm ($\sim 0.0042(16)$ R$_\odot$), which is consistent with one of an ONe WD with a mass of 1.15-1.34 M$_\odot$ \citep{kas19massveWD}. This WD mass is considerably more massive than the typical ones in CVs ($\langle M_{\rm WD} \rangle \approx 0.8 \pm 0.2 $ M$_\odot$; \cite{zor11SDSSCVWDmass,pal22WDinCVs}).

\section{Observations}
\label{sec:3}

\subsection{photometric observations}
\label{sec:3phot}

We used various telescopes and instruments for the optical--IR photometry of MASTER J0302, as summarized in table E1\footnote{Table E1 is available as supplementary data online.}. The log of our photometric observations is listed in table E3\footnote{Table E3 is available as supplementary data online.}.  All the observation epochs in this paper are described in the Barycentric Julian Date (BJD). 
The major part of our time-resolved photometric observations has been performed through the Variable Star Network Collaboration (VSNET; \cite{VSNET}), including the Fan Mountain Observatory\footnote{https://astronomy.as.virginia.edu/research/observatories/fan-mountain} and the Dark Sky Observatory\footnote{https://dso.appstate.edu/facilities}. A part of our multi-color observations was conducted in the framework of the Optical and Infrared Synergetic Telescopes for Education and Research (OISTER) collaboration, including 
Hiroshima Optical and Near-Infrared camera (HONIR) on the 1.5m Kanata telescope at  Higashi Hiroshima Observatory \citep{Aki14HONIR},
MITSuME on the 50cm telescope Akeno, the 50cm telescope Okayama, and the 105cm Murikabushi telescope at  Ishigakijima Astronomical Observatory \citep{kot05MITSUME1, Yat07MITSUME2, Shi08MITSUME3, yan10MITSUME4}, 
Nishiharima Infrared Camera (NIC) \citep{ish11nic, tak13nic} \footnote{http://www.nhao.jp/research/annual\_report/docs/ar2011-3.pdf} on the 2m Nayuta telescope at  Nishi-Harima Astronomical Observatory, 
Multi-wavelength SimultaneouS High throughput Imager and polarimeter (MuSaSHI) on the 55cm SACRA telescope at Saitama University Observatory \citep{Oas20MUSASHI}, and
TriCCS\footnote{http://www.o.kwasan.kyoto-u.ac.jp/inst/triccs/} on the 3.8m Seimei telescope at Kyoto University Okayama Observatory \citep{kur20seimei}.
We also made photometric observations at 2 frames per second with the wide-field CMOS camera Tomo-e Gozen \citep{tomoegozen} mounted on the 105 cm Schmidt telescope at the Kiso Observatory. The observations with Tomo-e Gozen were performed without a filter.

Our filtered observations are calibrated with the AAVSO standard stars TYC2-1228.1336.1 and TYC2-1228.1434. Their positions and magnitudes are listed in table E2 \footnote{Table E2 is available as supplementary data online.}. The zero point of the unfiltered data was adjusted to the $V$-band observations by the Kyoto University Observatory ($CV$ band).

We also obtained photometric time-domain catalog from the ASAS-SN Sky Patrol \citep{ASASSN, koc17ASASSNLC}, Zwicky Transient Facility (ZTF; \cite{ZTF}) alert broker Lasair \citep{lasair}, Gaia Photometric Science Alerts\footnote{http://gsaweb.ast.cam.ac.uk/alerts/alert/Gaia22age/}, and Asteroid Terrestrial-impact Last Alert System (ATLAS; \cite{ATLAS}) to examine the global light curve profile. The observations reported in CBAT and ATel \citep{zhi21J0302discovery, sar21J0302predicatel15081} are used to check the earliest light curve of this outburst as well.

The phase dispersion minimization (PDM;  \cite{PDM}) method was applied for the period analysis of superhumps in this paper.  The 90$\%$ confidence range of the $\theta$ statistics in the PDM method was determined following \citet{fer89error, pdot2}. Before period analysis, the global trend of the light curve was removed by subtracting a smoothed light curve obtained by locally weighted polynomial regression (LOWESS: \cite{LOWESS}).  Observed-minus-calculated ($O - C$) diagrams are presented for visualizing superhump period variations, which are sensitive to slight variations.  In this paper, we used 0.060281 d for the calculated ($C$).

\subsection{spectroscopic observations}
\label{sec:3sepc}

We performed the optical spectroscopic observations using the fiber-fed integral field spectrograph (KOOLS-IFU; \cite{mat19koolsifu}) mounted on the 3.8-m Seimei telescope \citep{kur20seimei}. Our observation log is summarized in table E5\footnote{Table E5 is available as supplementary data online.}.
We applied VPH-blue, VPH-red, VPH495, and VPH683 as a grism, which has a resolution of $R\sim 500$, 800, 1500, and 2000,  and wavelength coverage of 4100--8900 \AA, 5800--10200 \AA, 4300--5900 \AA, and 5800--8000 \AA, respectively.  The time-resolved spectroscopic observations were performed on BJD 2459547.22, 2459548.15, and 2459550.05 with VPH683, VPH495, and VPH683, and with the exposure times of 120, 120, and 180 s, respectively. Data reduction was performed using IRAF in the standard manner (bias subtraction, flat fielding, aperture determination, spectral extraction, wavelength calibration with arc lamps, and flux calibration with the standard star HD15318). Preliminary results have been reported in \citet{tag21J0302specatel15072,iso21J0302specatel15074}.

\section{Optical--IR Light curve}
\label{sec:4}

\subsection{Overall light curve}
\label{sec:4LC}

\begin{figure*}[tbp]
 \begin{center}
  \includegraphics[width=\linewidth]{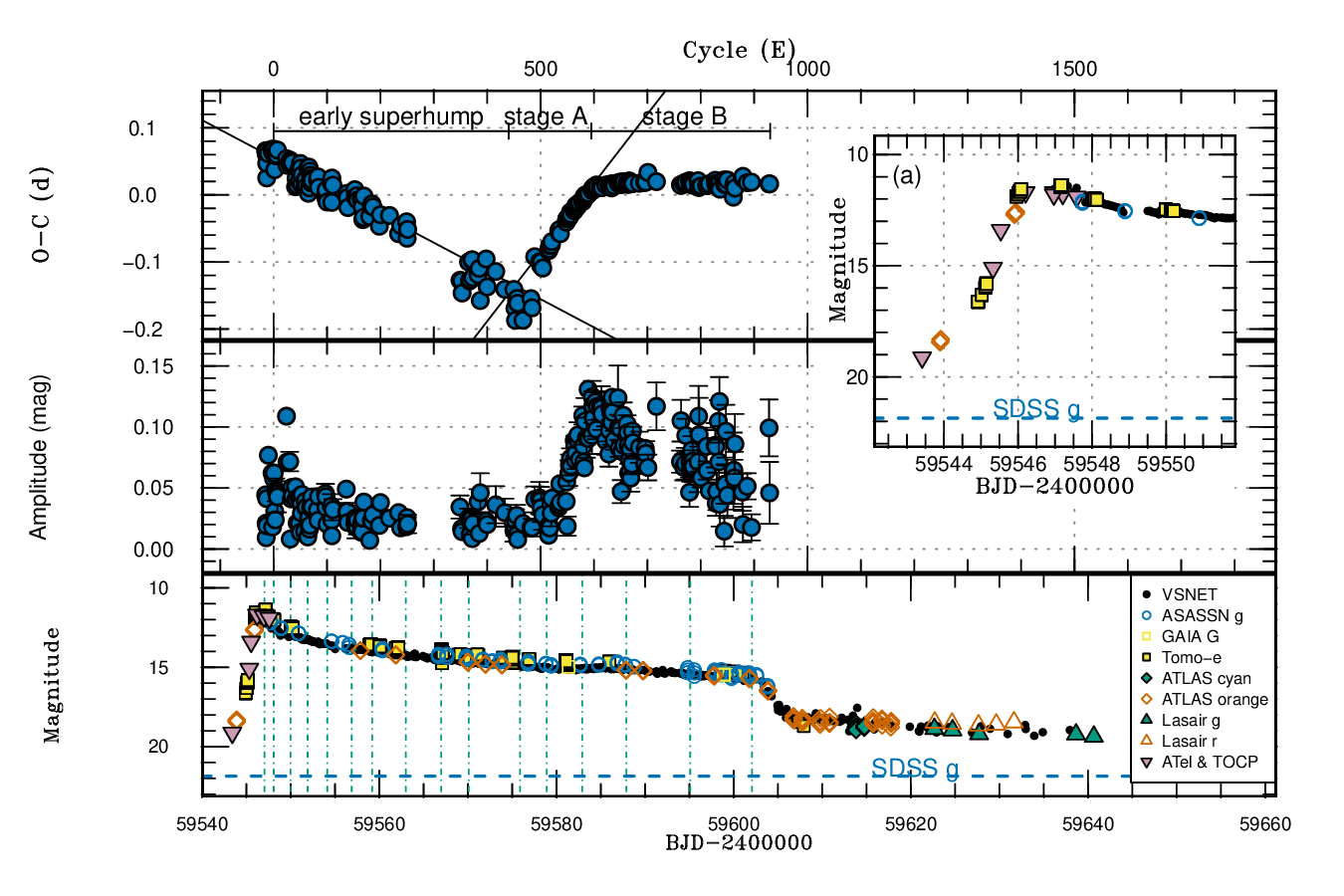}
 \end{center}
 \caption{Top panel: the $O - C$ diagram of superhump periods during the MASTER J0302 2021-2022 superoutburst. 
 0.060281 d was used for $C$. 
 The solid lines in the panel represent the early superhump period (0.05986 d) and the stage A superhump period (0.06131 d).
 Middle panel: the evolution of the superhump amplitudes in magnitude scale. 
 Bottom panel: the light curve of MASTER J0302 during the 2021 superoutburst.
 Refer to the legend for the representing dataset of each symbol.
 The VSNET data are binned in 0.06 d.
 The dot-dashed lines in the light curve show the epochs of our spectroscopic observations.
 The horizontal dashed line represents the quiescence magnitude in the SDSS $g$ band.
 The inserted panel (a) is the enlarged light curve around the outburst maximum.
 }
 \label{fig:2021OC}
\end{figure*}

The overall optical light curve of the MASTER J0302 outburst is presented in the lower panel of figure \ref{fig:2021OC}, along with the enlarged light curve around the outburst maximum in the inserted panel (a). MASTER J0302 reached its optical peak at 11.8 mag in the $V$ band on BJD 2459546-- 2459547. Its outburst amplitude is $\simeq 10.2$ mag, which is the largest in WZ Sge-type DNe and, therefore, in DNe regardless of subtypes \citep{kat15wzsge, tam20j2104}. Furthermore, this outburst entered a rapid decline phase on BJD 2459604.0, giving its duration as $\simeq$ 60 d. This is also the longest in WZ Sge-type DN outbursts to our best knowledge (see Section \ref{sec:72duration}).

Its outburst rise took more than 3.0 d and the rise timescale is 0.31(1) d mag$^{-1}$ via the linear regression of the light curve. This rise timescale is comparable to that of other WZ Sge-type DNe \citep{otu16DNstats}. The decline timescale during BJD 2459547.5 -- 2459556.0, 2459556.0 -- 2459580.0 and 2459585.0 -- 2459603.0 is 5.088 (7), 17.99(2), and 20.44(5) d mag$^{-1}$, respectively. The outburst stopped fading and showed a slight brightening around BJD 2459580-- 2459585, as commonly observed in  WZ Sge-type DNe \citep{can01wzsge,osa03DNoutburst, kat15wzsge}. After the rapid decline of the outburst, MASTER J0302 showed steady fading with a decline timescale 17.0(1) d mag$^{-1}$. No rebrightening was observed.

\subsection{Superhump evolution}
\label{sec:4SH}

\begin{figure}[tbp]
 \begin{center}
  \includegraphics[width=\linewidth]{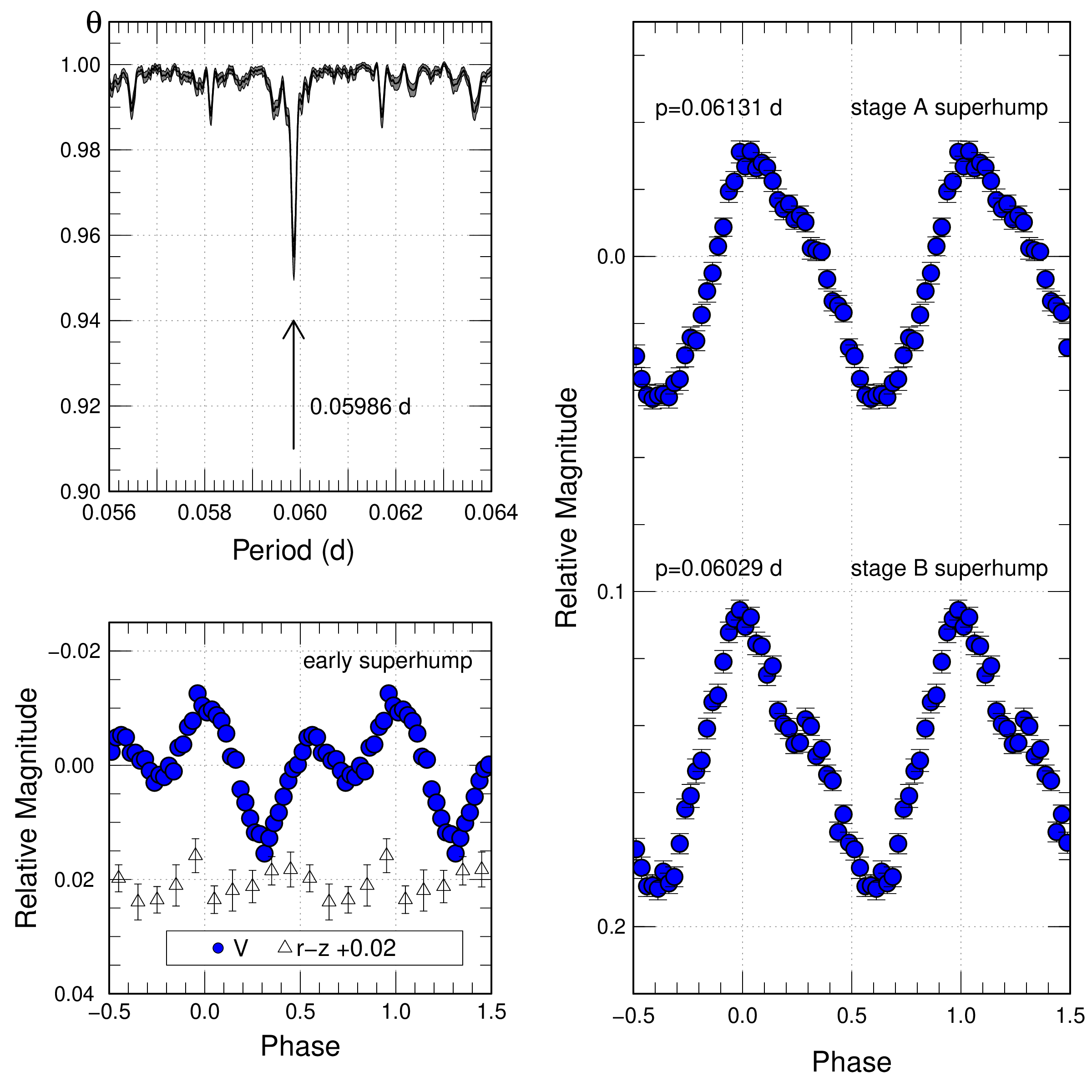}
 \end{center}
 \caption{
 Upper left-handed panel: $\theta$-diagram of the PDM analysis using data on BJD 2459549.8 -- 2459563.5 (early superhump phase).
 The gray area represents the 90 $\%$ confidence range of $\theta$ statistics by the PDM method. 
 Lower left-handed panel: phase-averaged profile of early superhumps (filled circles) along with the $r-z$ color (open triangles) on BJD 2459551.1 observed with SaCRA.
 Right-handed panel: phase-averaged profiles of stage A (upper) and stage B superhumps (lower).
}
 \label{fig:shs}
\end{figure}

Our intensive time-resolved observations have detected superhumps throughout the superoutburst. The top panel of figure \ref{fig:2021OC} shows the $O-C$ diagram of superhump periods using $C = 0.060281$ d, and the middle panel represents the amplitude of superhumps on the magnitude scale. The times of the superhump maxima are presented in table E4 \footnote{Table E4 is available as supplementary data online.}. Based on the $O - C$ diagram and amplitude evolution, we determined the early superhump, stage A ordinary superhump, and stage B ordinary superhump phases as BJD 2459543.4-- 2459575.2, 2459575.2-- 2459584.7, and 2459584.7-- 2459604.0, respectively. Due to the lack of time-resolved data after the rapid decline, late-stage superhumps were not resolved.

In figure \ref{fig:shs}, the results of the PDM analysis (upper left-handed panel) and phase-averaged profile (lower left-handed panel) of early superhumps, together with the phase-averaged profiles of stage A and stage B superhumps (right-handed panel) are presented. Superhump periods are determined as 0.05986(1), 0.06131(2), and 0.060286(6) d for early superhumps, stage A superhumps, and stage B superhumps, respectively.
The early superhumps show a double-waved profile with $\simeq 0.03$ mag amplitude. In figure E1\footnote{Figure E1 is available as supplementary data online.}, we compare the phase-averaged early superhump profiles of MASTER OT J030227.28+191754.5 (this paper), MASTER OT J005740.99$+$443101.5, WZ Sge, OT J012059.6$+$325545, and V466 And \citep{kat15wzsge}. The early superhump profile of MASTER J0302 well resembles that of other WZ Sge-type DNe.
The $r-z$ color of early superhumps on BJD 2459551.1 (open triangles in lower left-handed panel of figure \ref{fig:shs}) does not show significant change above an error range 0.01 mag throughout the whole orbital phase. These early superhump features are all consistent with those of other WZ Sge-type DNe. We hence treated the early superhump period as the orbital period in the rest of this paper following the usual manner in WZ Sge-type DNe.

Using the periods of the early and stage A superhump and applying the relation by \citet{kat13qfromstageA, kat22updatedSHAmethod}, the mass ratio $q$ of MASTER J0302 is obtained as 0.063(1). The observed behaviors of superhumps in MASTER J0302 are consistent with those of other low mass-ratio DNe and the expectations in the TTI model \citep{kim18asassn16dt16hg};
small amplitude of ordinary superhumps ($\leq$ 0.08 mag), 
long-lasting stage A superhumps ($\approx$9.5 d),
small $P_{\rm dot}$ during stage B superhumps ($-$0.3(3) $\times 10^{-5}$),
a large decrease in the superhump period between the stage A and stage B superhump transition (1.7(1)\%), 
and a long decline timescale in the plateau stage with ordinary superhumps (20.44(5) d mag$^{-1}$).
Therefore, although the overall light curve is exceptionally energetic, the superhump behaviors in MASTER J0302 reasonably align with the TTI model.

We note that, except for the superhump periods, we did not detect any other coherent variability (i.e. spin period of the primary WD), confirming the non-detection of spin periods in X-ray \citep{kim23j0302}. Therefore, the magnetic activity of the primary WD should be low enough (typically $<10^5$ G; \cite{war95book}) that pole accretion does not take place during this outburst.

\subsection{color evolution}
\label{sec:4col}

\begin{figure*}[tbp]
 \begin{center}
  \includegraphics[width=\linewidth]{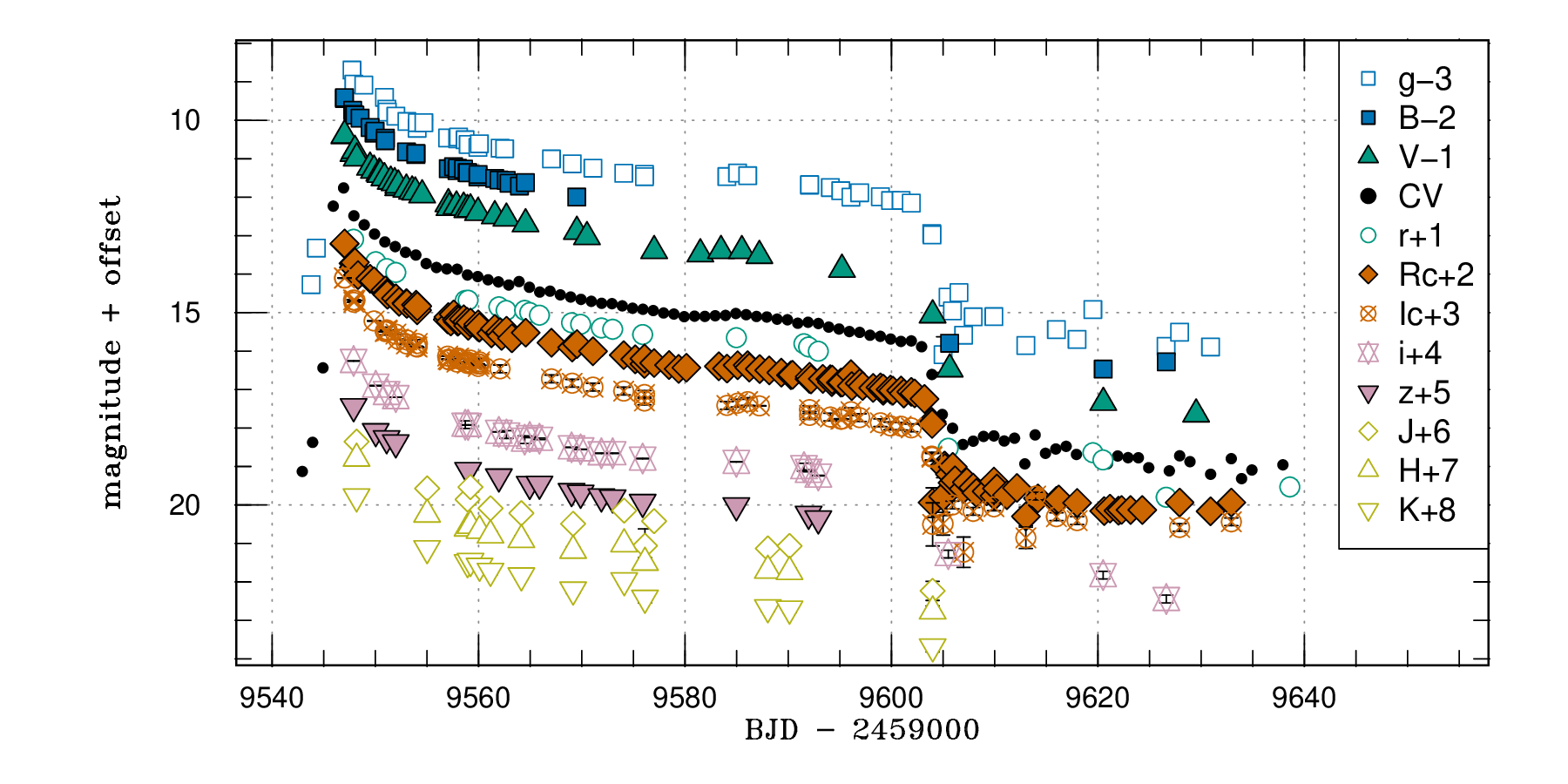}\\
  \includegraphics[width=0.45\linewidth]{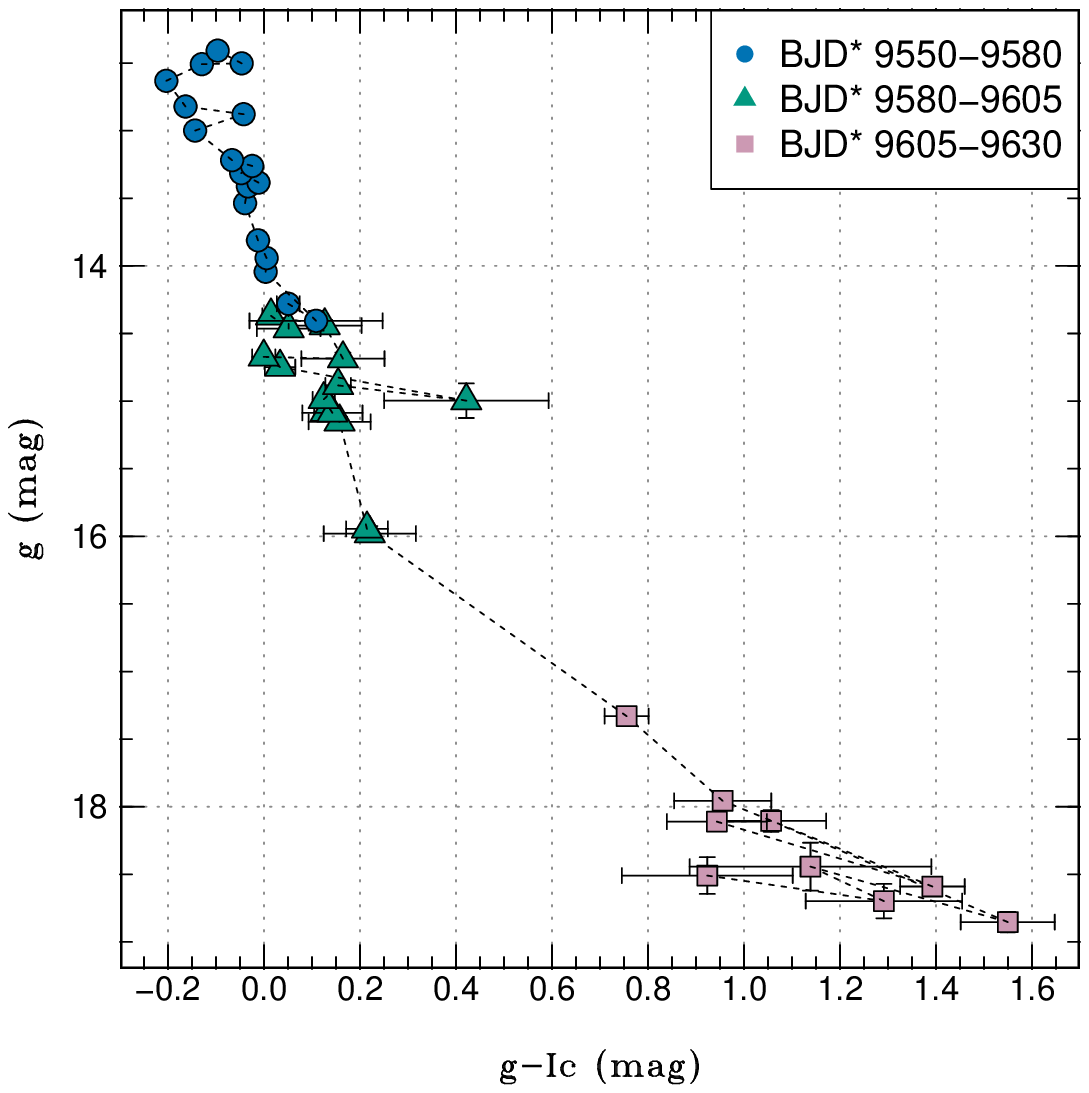}
  \includegraphics[width=0.45\linewidth]{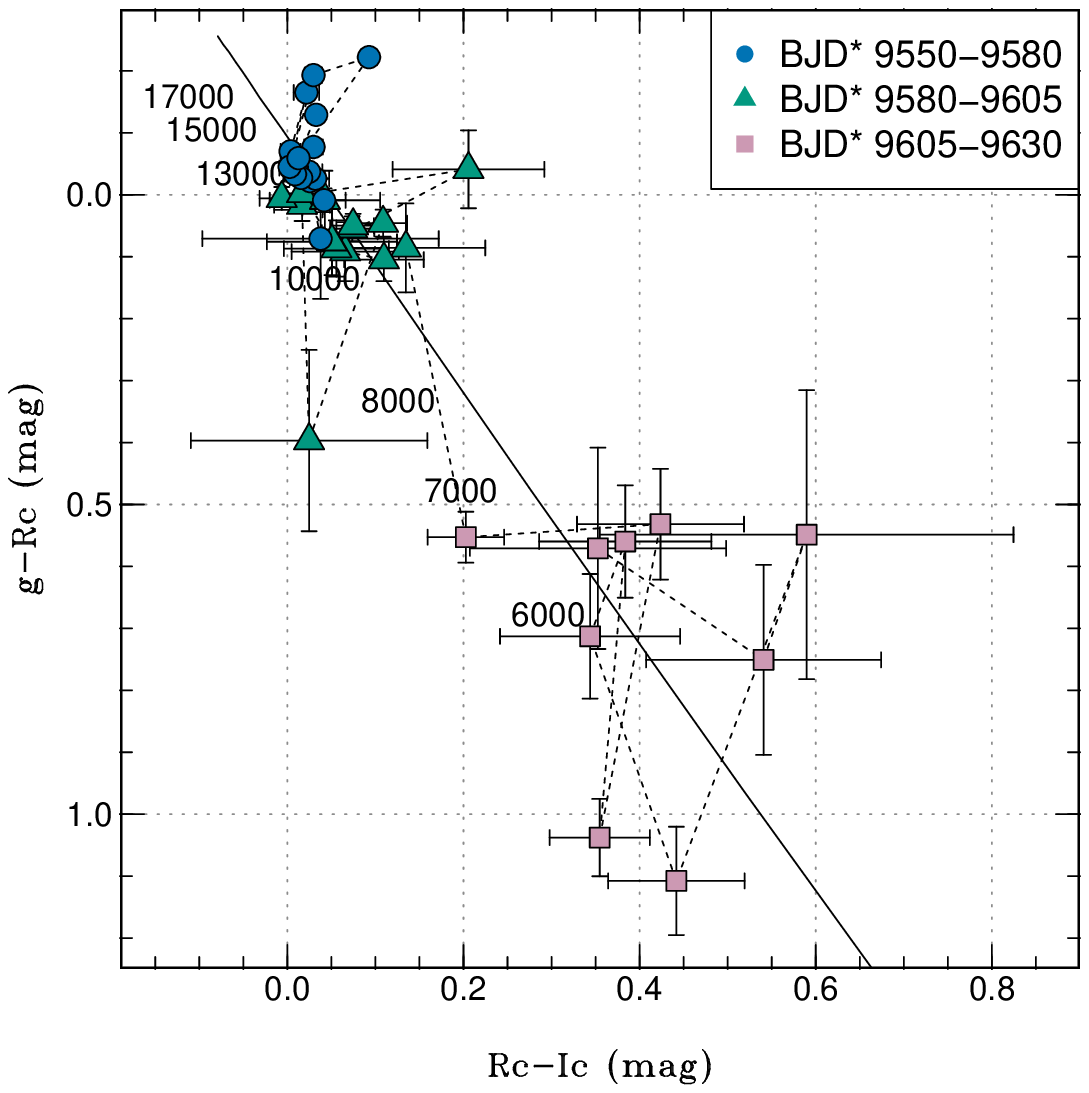}
 \end{center}
 \caption{Top panel: the multi-color light curves of MASTER J0302 in 
 $g$,  $B$, $V$,  $CV$,  $r$,  $R_{c}$,  $I_{c}$, $i$,  $z$,  $J$,  $H$,  and $K_{s}$ bands from upper to lower.
 The light curves are vertically shifted for better visualization, with the shifted magnitudes shown in the figure legend.
 $CV$ data are binned in 1 d.
 Bottom panels: magnitude-color diagram in $g$ vs $g-I_{\rm c}$ band (left-handed panel) and color-color diagram in $g, R_{\rm c}, I_{\rm c}$ bands (right-handed panel).
 The solid line in the color-color diagram represents color of blackbody emission with the corresponding temperature shown on the left side.
 The different symbols represent different phases of the outburst; BJD 2459550.0 -- 2459580.0, 2459580.0 -- 2459605.0, and 2459605.0 -- 2459630.0 with the blue circles, green triangles, and pink squares, respectively.}
\label{fig:colcolmap}
\end{figure*}

In figure \ref{fig:colcolmap}, the multi-color light curves from optical to IR (upper panel), and the evolution in magnitude-color (lower left panel) and color-color (lower right panel) diagrams. All these panels present the extinction-corrected results. In the lower panels, the observations with MITSuME ($g, R_{\rm c}, I_{\rm c}$ bands) are presented. The different symbols in the lower panels represent the different phases of the outburst; the early superhump phase BJD 2459550.0 -- 2459580.0 (blue circles), the ordinary superhump phase BJD 2459580.0 -- 2459605.0 (green triangles), and the phase after the rapid decline BJD 2459605.0 -- 2459630.0 (pink squares). The color-color map also presents colors of blackbody emission with the solid line. The color temperature largely corresponds to that of the outer disk, which contributes most of the optical continuum \citep{may80ADspec, ech83DNphotometry}. Figure E2\footnote{Figure E2 is available as supplementary data online.} presents the change of the MITSuME $g-I_{\rm c}$ color along the observation epochs.

All the light curves from optical to IR essentially follow the $V$ band light curve. Around the outburst maximum the $g-I_{\rm c}$ color is around zero, and the corresponding color temperature was 15000 K. As the outburst declined, the color reddened $\sim 0.1$ mag, and the temperature cooled to 11000 K. Other than the gradual reddening across the outburst, we did not find any significant change of color around the superhump stage transition. After the rapid decline, the color became much redder ($g-I_{\rm c}$ $\sim$ 1.0) and the temperature dropped $\sim$ 6000 K, reflecting the end of the outburst. These colors and temperatures, and their evolutions, are within the range of those observed in other DN outbursts [\citet{mat09v455and, shu21aylac} and the references therein].

\section{Spectral evolution}
\label{sec:5}

Our spectroscopic observations covered the entire phase of the outburst, from the outburst maximum to just before the rapid decline. The dot-dashed lines in the bottom panel of figure \ref{fig:2021OC} show their epochs. Figures \ref{fig:specevolline} and E2\footnote{Figure E3 is available as supplementary data online.} present the evolution of normalized spectra throughout the outburst. In table E6\footnote{Table E6 is available as supplementary data online.}, we summarize the equivalent width (EW) of our interested lines. We note that, since the EW is defined for absorption lines, a negative EW means an emission line. In the following subsections, we first present the spectral evolution across the superoutburst in section \ref{sec:5outburst}, and then our time-resolved spectra around the outburst maximum in section \ref{sec:5atmax}.

\subsection{Spectral evolution across the superoutburst}
\label{sec:5outburst}

\begin{figure*}[tbp]
 \begin{center}
  \includegraphics[width=0.45\linewidth]{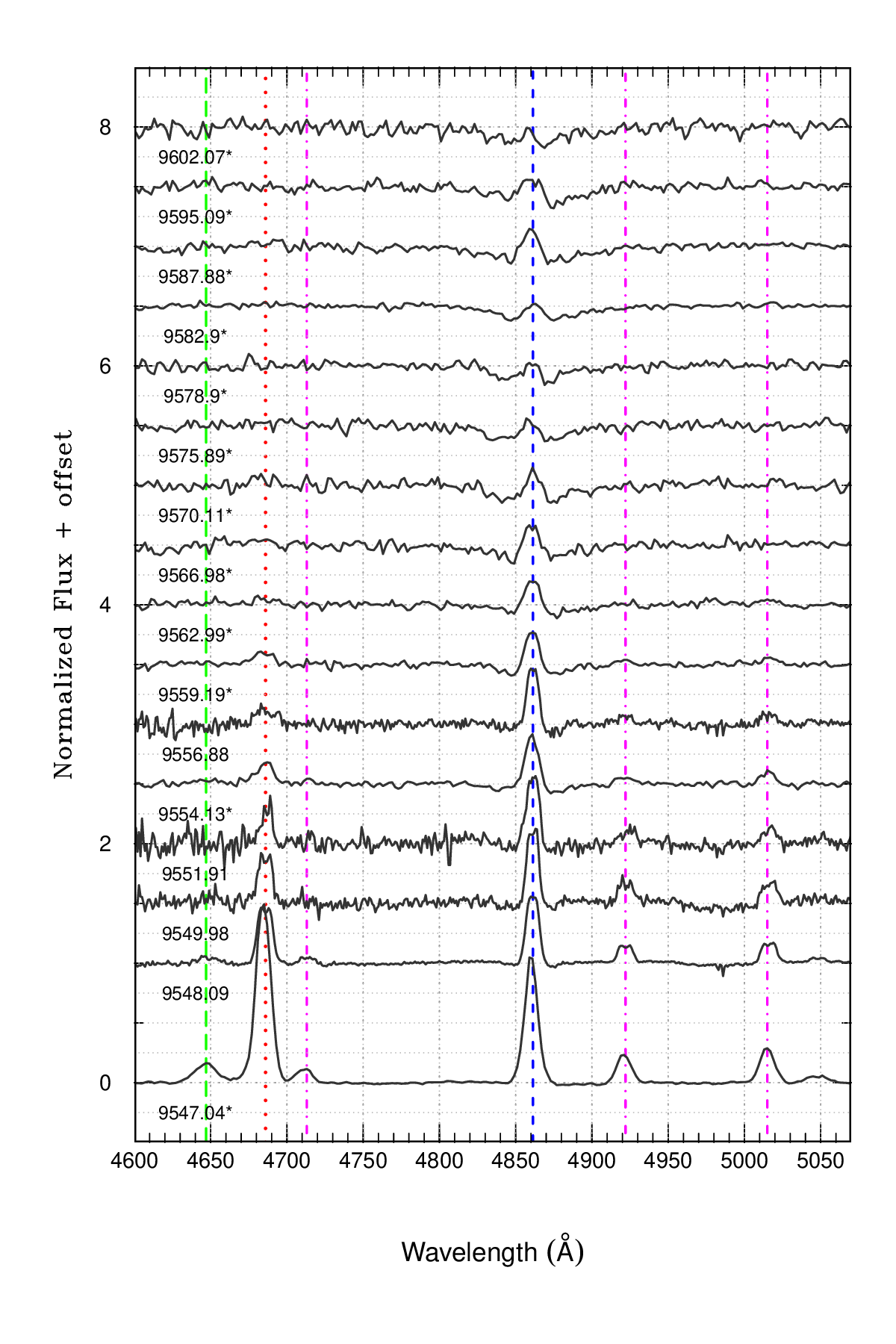}
  \includegraphics[width=0.45\linewidth]{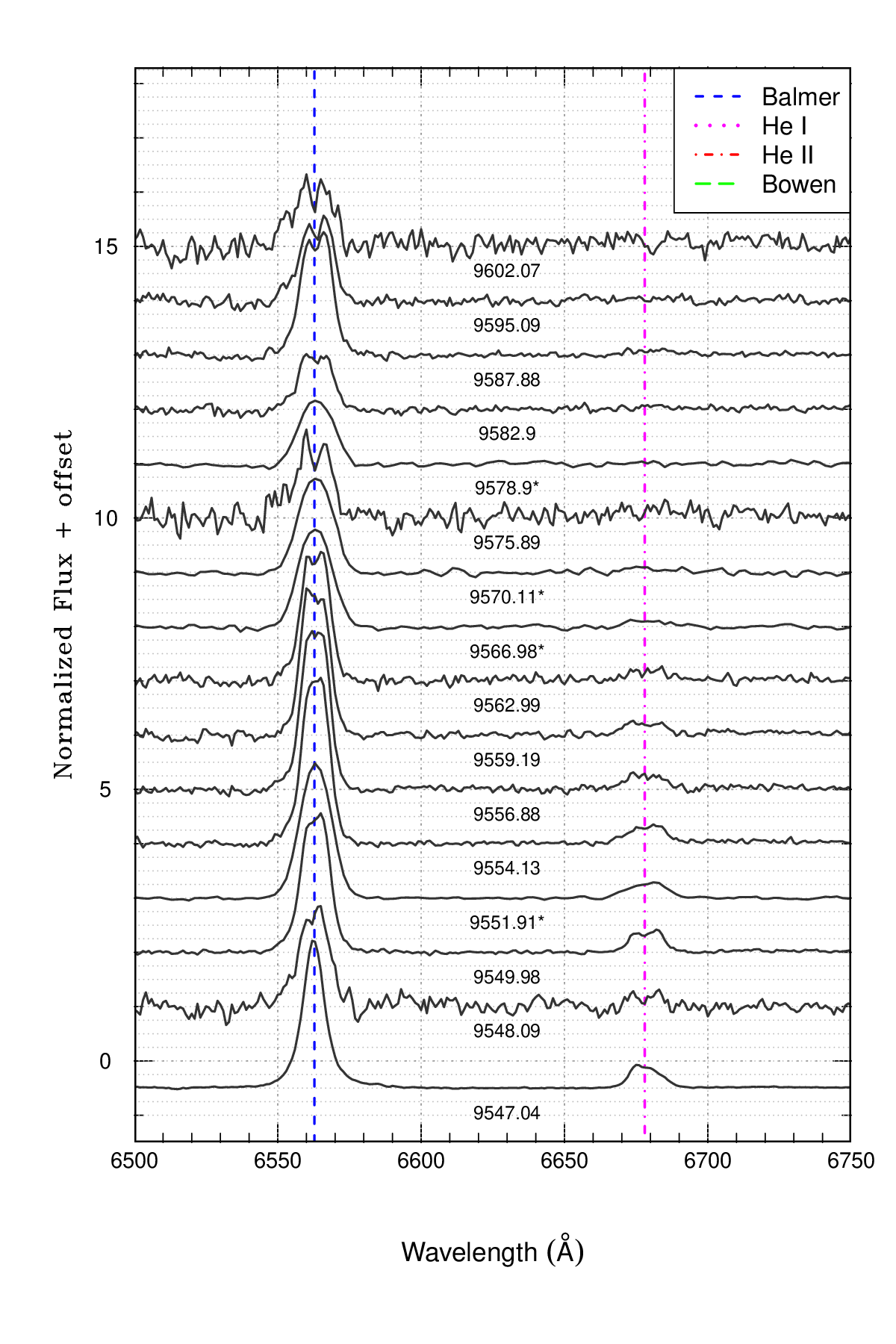}
 \end{center}
 \caption{Time evolution of the normalized spectra from lower to upper.
 The left-handed and right-handed panels cover 4600--5070 and 6500--6750 \AA, respectively.
 The dashed, dot-dashed, dotted, and long-dashed lines represent the wavelength of Balmer, He~\textsc{i}, He~\textsc{ii}, and Bowen blends, respectively.
 The epochs (BJD - 2450000) of our observations are shown below the spectra.
 Epochs with an asterisk were observed with VPH-blue, while epochs without an asterisk were observed with VPH683 or VPH495 with a better wavelength resolution. }
 \label{fig:specevolline}
\end{figure*}

\begin{figure*}[tbp]
 \begin{center}
  \includegraphics[width=\linewidth]{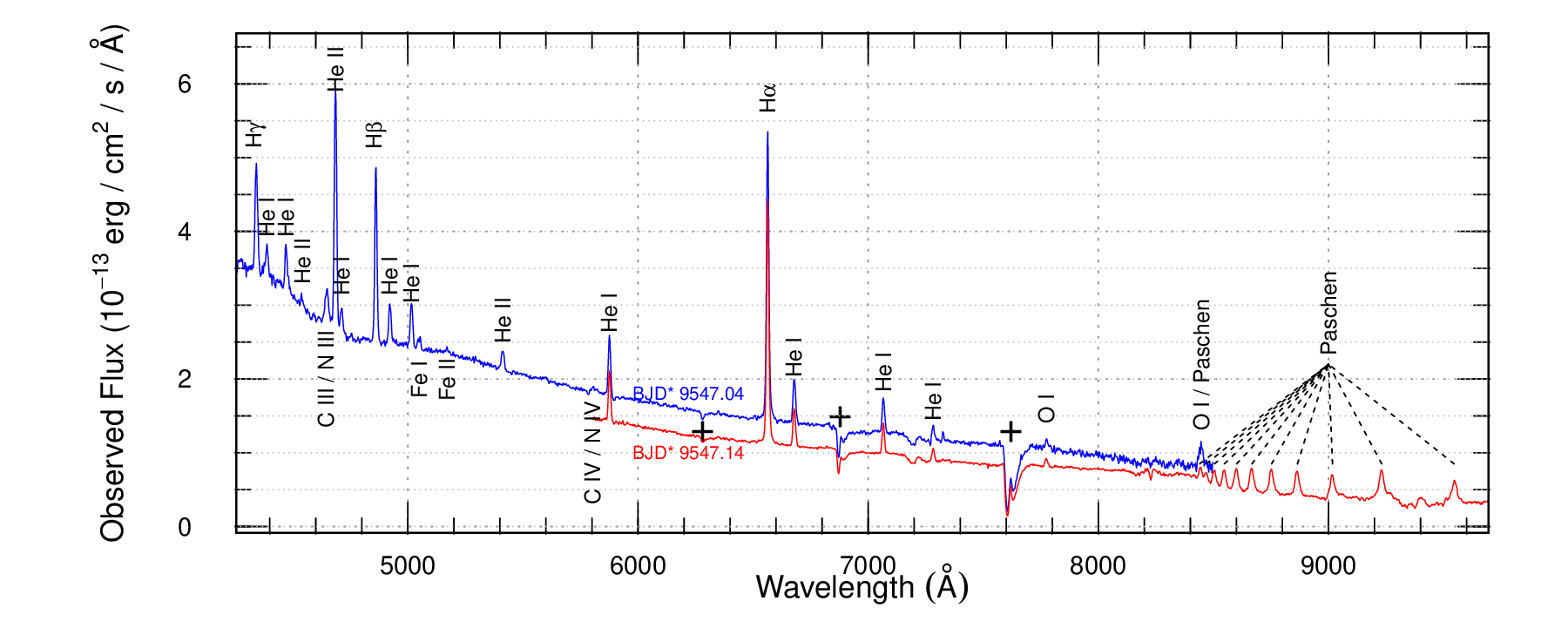}
 \end{center}
 \caption{Flux-calibrated spectra on BJD 2459547.04 with VPH-blue (upper) and on BJD 2459547.14 with VPH-red (lower).
 Telluric absorption lines are marked with $+$.}
 \label{fig:earlyspec}
\end{figure*}

Our first spectra were observed on BJD 2459547. The flux-calibrated spectra are presented in figure \ref{fig:earlyspec}. These spectra show a blue continuum attributing to a multi-temperature disk blackbody, together with numerous emission lines of Balmer, Paschen, He~\textsc{i} 4388, 4471, 4713, 4922, 5015, 5876, 6678, 7065, He~\textsc{ii} 4542, 4686, 5421, C~\textsc{iii} / N~\textsc{iii} Bowen blend, C~\textsc{iv} / N~\textsc{iv} blend, Fe~\textsc{i} 5049, Fe~\textsc{ii} 5169, and O~\textsc{i} 7773, 8446. The EWs of H$\alpha$, H$\beta$, and He~\textsc{ii} 4686 were $-$46.7(4), $-$11.2(3), and $-$18.7(2) \AA, respectively.
In the mid-resolution spectrum, H$\alpha$ showed a single-peaked profile with its typical full line width narrower than 500 km s$^{-1}$, while He~\textsc{i} 6678 gives a hint of a double-peaked profile. In terms of an EW, He~\textsc{ii} 4686 showed stronger emission than H$\beta$.
These points, (1) strong high-excitation lines and (2) narrow-peaked emission lines remind the spectra of a DN V455 And around the outburst maximum \citep{tam21seimeiCVspec,tam22v455andspec}.

One day later, on  BJD 2459548.09, the emission lines became much weaker, especially for the high-excitation lines, while their EWs ($-$23.8(3), $-$4.1(1), $-$6.1(1) \AA~ for H$\alpha$, H$\beta$, and He~\textsc{ii} 4686, respectively) indicate still stronger emission-line profiles in MASTER J0302 than those in other WZ Sge-type DNe \citep{tam21seimeiCVspec}. Since this epoch, all emission lines including H$\alpha$ clearly showed double-peaked profiles in the middle-resolution spectra. The peak separations of H$\alpha$, He~\textsc{i} 6678, and He~\textsc{ii} 4686 were $\sim$ 200, 400, and 380  km s$^{-1}$, respectively, on BJD 2459548.09. The typical full-width half max (FWHM) of these lines was $\lesssim$ 800 km s$^{-1}$.

H$\gamma$ became absorption lines after BJD 2459559.19. An absorption component also began to accompany H$\beta$ from BJD 2459548.09, and the net EW became positive after BJD 2459566.98. He~\textsc{ii} 4542, He~\textsc{ii} 5411, and C~\textsc{iv} / N~\textsc{iv} blend became unrecognized in our spectrum on BJD 2459548.09, 2459549.99, and 2459549.99, respectively. He~\textsc{ii} 4686 and Bowen blend are not recognized after BJD 2459562.99 and BJD 2459556.88, respectively, which correspond to the middle of the early superhump phase. After BJD 2459582.90, the peak separation of H$\alpha$ expanded to $\geq$ 300 km s$^{-1}$.

The emission components of Balmer and He~\textsc{i} lines once again became stronger on BJD 2459587.89. The absolute value of the H$\alpha$ EW changed by $\sim$ 70 \% (see table E6). This epoch corresponds to the appearance of stage B superhumps; however, it was later than the slight brightening in the $V$-band light curve around BJD 2459580 -- 2459585, which corresponds to the later stage of stage A superhumps. 
In the TTI model, stage A superhumps correspond to the excitation of tidal instability at the 3:1 resonance radius, and then superhumps switch to stage B because of inner propagation of eccentricity \citep{Pdot, nii21a18ey}. Based on this view, the time delay between the light curve and the EW evolution can be understood as the continuum emission brightened due to the increase of the local accretion rate at the outer disk (stage A superhump), and the line emission excited due to the increased UV/X-ray irradiating photons coming from the inner disk (stage B superhump).

\subsection{Time-resolved spectra around the outburst maximum}
\label{sec:5atmax}

\begin{figure*}[tbp]
 \begin{center}
  \includegraphics[width=\linewidth]{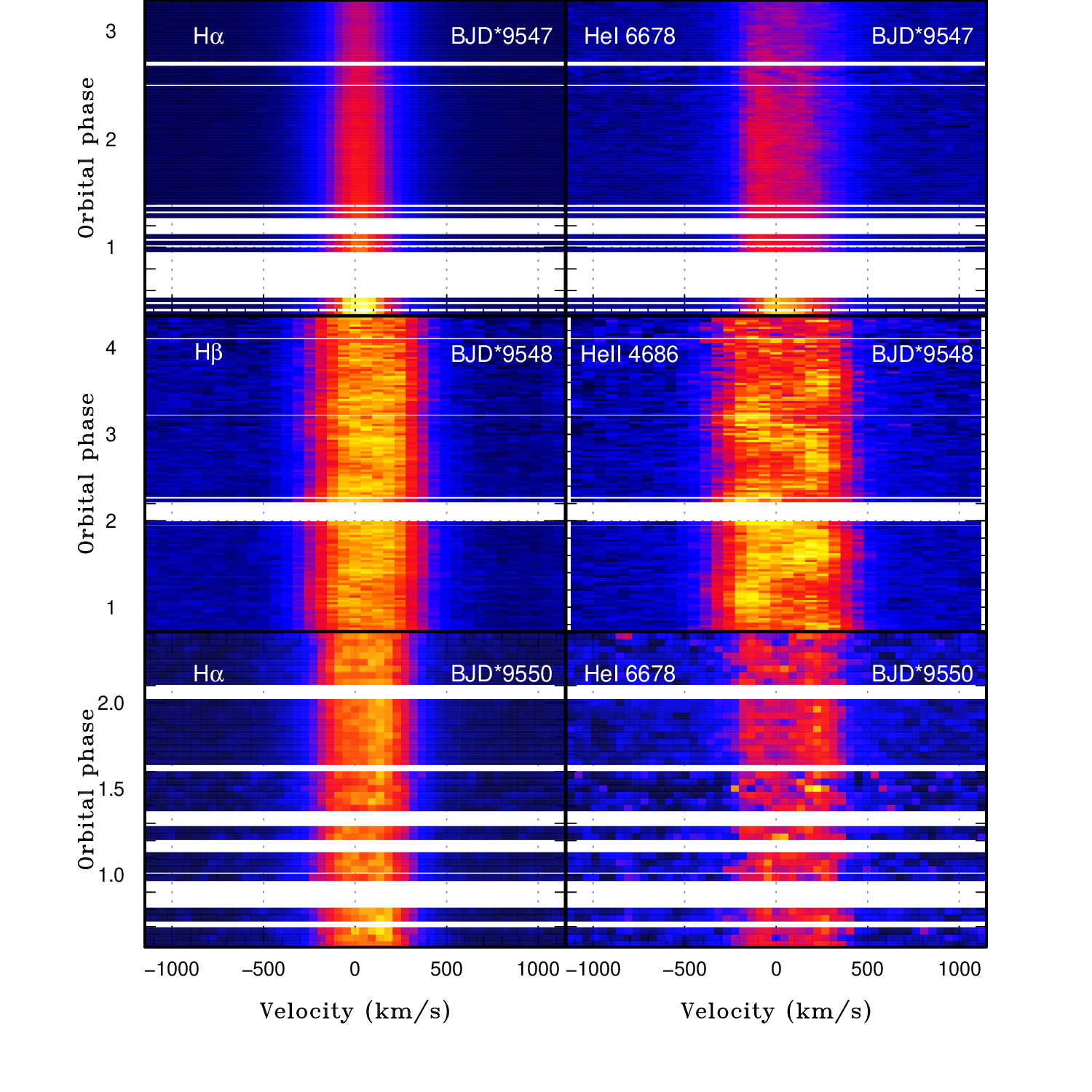}
 \end{center}
 \caption{Time-resolved dynamic spectra in the orbital phase--velocity plane on BJD 2459547.2, 2459548.2, and 2459550.1 from upper to lower. 
 The color scale shows the flux intensity normalized by the continuum.
 Data on BJD 2459547.2 and 2459550.1 cover H$\alpha$ and He~\textsc{i} 6678, 
 and that on BJD 2459548.2 cover H$\beta$ and He~\textsc{ii} 4686.
 The blank rows are due to bad weather or dead time of the telescope.
 The orbital phase of the spectra was determined following the Appendix A1.
}
 \label{fig:phasespec}
\end{figure*}

\begin{figure}[tbp]
 \begin{center}
  \includegraphics[width=\linewidth]{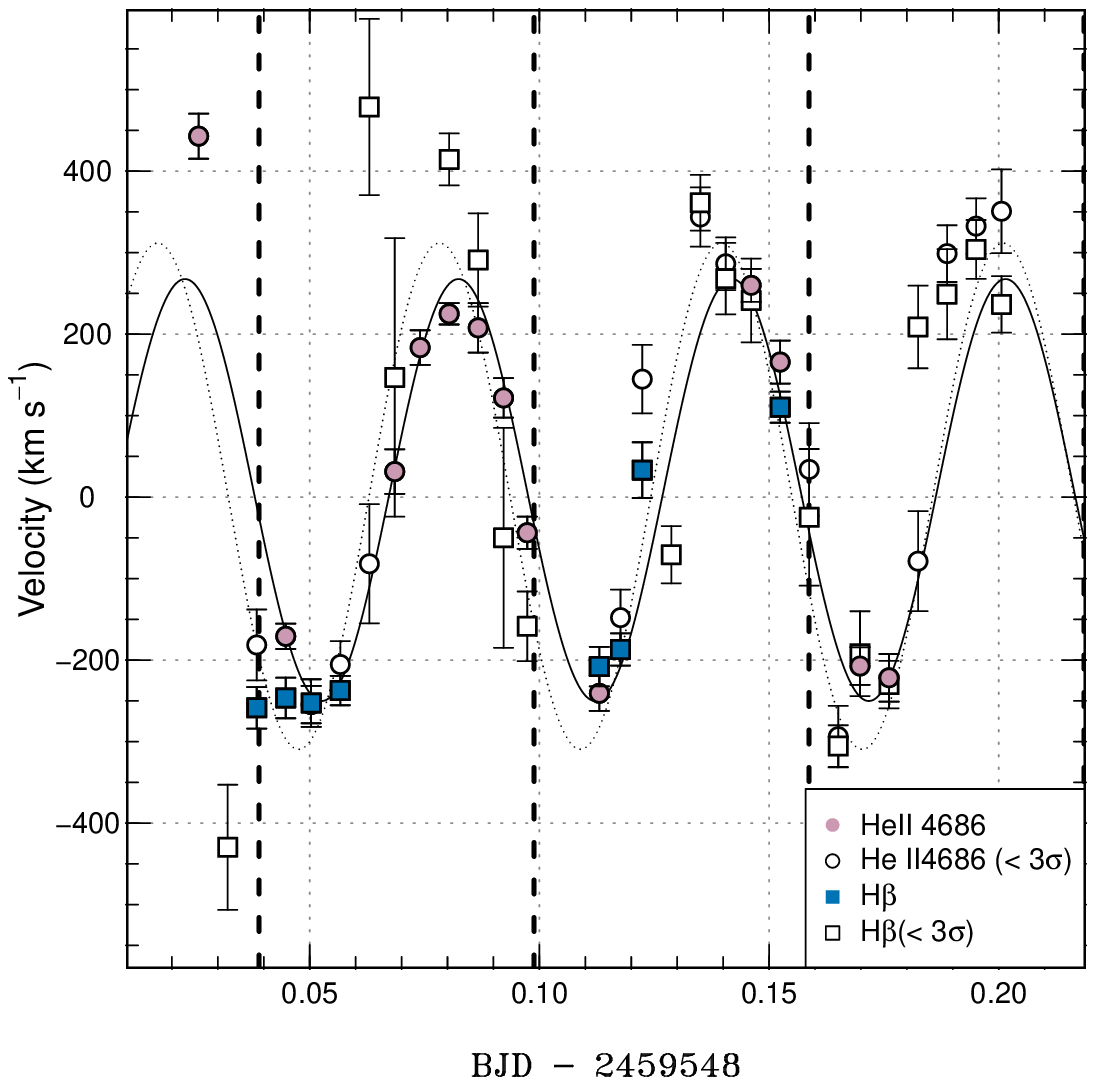}
 \end{center}
 \caption{
 Radial velocities of H$\beta$ (squares) and He~\textsc{ii}~4686 (circles) on BJD 2459548. 
 Each point is binned in 0.006 d ($\sim 0.1$ orbital phase).
 The filled and open symbols represent those detected above and below 3 $\sigma$ level, respectively.
 The dotted and solid curves represent the best-fit sine curve for these qualified radial velocities of H$\beta$ and He~\textsc{ii}~4686.
 The vertical dashed lines correspond to the phase zero following equation \ref{eq:ephemeris}.
}
 \label{fig:radvel}
\end{figure}

Figure \ref{fig:phasespec} presents our time-resolved spectra around the outburst maximum. The observations on BJD 2459547.2 and 2459550.1 covered H$\alpha$ and He~\textsc{i} 6678, whereas those on  BJD 2459548.2 did H$\beta$ and He~\textsc{ii} 4686. The orbital phase of the spectra in figure \ref{fig:phasespec} follows equation \ref{eq:ephemeris}.

On BJD 2459547.2, one can see a change of EWs of both H$\alpha$ and He~\textsc{i} 6678 even during our observation run. Radial velocities of the entire line profiles were not detected in either of our time-resolved spectra. We did not find any phase-dependent change of the line profiles in H$\alpha$ and He~\textsc{i} 6678 on both nights. Conversely, an S-wave-like structure in He~\textsc{ii} 4686 and marginally in H$\beta$, possibly modulating with the orbital period for three cycles, can be recognized.

We hereby check the radial velocity of this S-wave component in H$\beta$ and He~\textsc{ii} 4686. Unfortunately the signal-to-noise ratio of the Bowen blend was not high enough for these analyses.
First, we subtracted the averaged line profile of this night from each binned one in 0.006 d ($\approx$ 0.1 orbital phase). Then we measured the central wavelength of the residual by fitting it with Gaussian function. Figure \ref{fig:radvel} shows the resulting radial velocities of the S-wave component in H$\beta$ (square) and He~\textsc{ii} 4686 (circle). The filled symbols represent the epochs whose amplitude of the fitted Gaussian function exceeds the three-sigma level of the standard deviation of the residual, whereas the open symbols represent those falling this criterion. Finally, we fitted the former qualified ones with the sine curve (equation \ref{eq:RVsin}) as a function of time $T$, with free parameters of radial velocity amplitude $K$, phase zero epoch $T_0$, period $P$, and systematic velocity $\gamma$.
\begin{equation}
    \label{eq:RVsin}
    V(T) = K \sin \left(\frac{2 \pi \left( T - T_0\right)}{P} \right) + \gamma
\end{equation}
By fitting H$\beta$ and He~\textsc{ii} 4686 individually, the yielded radial velocity amplitude ($310 \pm 89$ and $259 \pm 21$ km s$^{-1}$ for H$\beta$ and He~\textsc{ii} 4686, respectively), phase zero epoch (BJD 2459548.002(3) and 2459548.008(2)), and period (0.061(2) and 0.060(2)), are consistent between H$\beta$ and He \textsc{ii}~4686 within three sigma level. Both derived periods are also consistent with the early superhump period.

\section{Analysis}
\label{sec:6}

\subsection{Binary parameters of MASTER J0302}
\label{sec:6params}

Based on the observational results described in the previous sections, we derive some important binary parameters of MASTER J0302. Here we introduce two different cases for adopting that
(1) superhumps in MASTER J0302 follow the same manner as other WZ Sge-type DNe (TTI case) and 
(2) the S-wave component observed in H$\beta$ and He~\textsc{ii} 4686 originates from a (irradiated) secondary (irradiated secondary case).
We list them in table \ref{tab:1}. In either case, we adopted the orbital period (0.05986(1) d) and the mass ratio (0.063(1)), derived from the observations of superhumps, as these methods have been tested in various types of CVs and compact binaries (see \cite{kat22updatedSHAmethod} and references therein).  Its inclination is limited as $\lesssim 70^\circ$, based on the non-detection of eclipses.
It should be acknowledged that these two cases are introduced only to find the binary parameters of MASTER J0302. Irradiation phenomenon itself on the secondary star generally happens in CVs (see \cite{osa03DNoutburst, osa04EMT, via07irrsecondary} for WZ Sge-type DN cases). We note that in passing future dynamical studies will be needed to confirm these binary parameters, as either case relies on the outburst physics in DNe.

\begin{table*}
    \caption{The binary parameters of MASTER J0302 under the TTI and irradiated secondary (Irr. sec.) case}
    \label{tab:1}
    \centering
    \begin{tabular}{cccccccc}
    \hline
    case    & orbital period & mass ratio & phase zero & inclination & WD mass & WD radius & distance \\
                & $P_{\rm orb}$ [d] & $q$ & $T_{0}$ [BJD - 2450000] & $i$ [$^\circ$] & $M_{\rm WD}$ [$M_\odot$]& $R_{\rm WD}$ [0.01 $R_\odot$] & $D$ [pc] \\
    \hline
    TTI & 0.05986(1) & 0.063(1) & 9548.0391$^{+0.0024}_{-0.0021}$ & 60$\pm{10}^\circ$ & 1.25$^{+0.09}_{-0.10}$ & 0.42$\pm{0.16}$ & 685$^{+85}_{-80}$\\
    Irr. sec. & 0.05986(1) & 0.063(1) & 9548.002(3) & \multicolumn{2}{c}{see figure \ref{fig:rvtoincl}} & -- & -- \\
    \hline
    \end{tabular}
\end{table*}

\subsubsection{TTI case}
\label{sec6:ttiparam}

Based on the TTI model explaining DN superoutbursts, various relations between superhump properties and binary parameters have been established. Although these have been fairly tested in ordinary DNe, there might be some limitations in applying them in unique systems. In the case of MASTER J0302, as discussed in section \ref{sec:4}, despite its 60-d long and 10.2-mag amplitude outburst, its superhump properties are reasonably consistent with other low mass-ratio DNe and the TTI model. This justifies its binary parameters in this case.

To determine the phase zero timings of the orbital ephemeris (i.e., inferior conjunction of the secondary), we compared the early superhump profile of MASTER J0302 with those of the other WZ Sge-type DNe whose orbital ephemeris has been studied in detail. We summarize our method and error estimation with the Bayesian technique in Appendix A1. \footnote{This technique may give a half-phase shift due to the double-peaked profile of early superhumps.} The phase zero epoch $T_0$ is obtained as BJD 2459548.0391$^{+0.0024}_{-0.0021}$ (table \ref{tab:1}). \citet{kat22WZSgecandle} formulated the relation between inclinations and early superhump amplitudes by interpreting that the scatter of the absolute magnitude at the appearance of ordinary superhumps is due to limb darkening effect.
According to this empirical relation, detection of early superhumps limits the inclination larger than $40^\circ$, and 0.03 mag amplitude of early superhumps corresponds to the inclination $i = 60^\circ$. We conservatively estimate its error as 10$^\circ$ following \citet{kat22WZSgecandle}.
\citet{kim23j0302} estimated the WD radius as $0.0042(16)~R_{\odot}$ and mass as 1.15--1.34 M$_\odot$ based on the X-ray observations in outburst. Moreover, \citet{kat22WZSgecandle} estimated its distance to be 720 pc by treating the appearance of ordinary superhumps (m$_V \sim 14.9$) as a standard candle. However, they did not consider the Galactic extinction and the effect of an intrinsically larger disk in a more massive WD system (see equation 11 in \cite{war87CVabsmag}). Applying these corrections, its distance is updated as 670--700 pc, depending on the WD mass, still within the error range in \citet{kat22WZSgecandle}.  Its typical error is 10$\%$ based on the uncertainty of the optical magnitude at the appearance of ordinary superhumps \citep{kat22WZSgecandle}.

According to this distance and the Galactic extinction, the $V$-band absolute magnitude of MASTER J0302 at outburst maximum is $M_{\rm V} \sim $ 2.4--2.2 mag. This is the brightest in known WZ Sge-type DNe \citep{tam20j2104,kat22WZSgecandle}, and one of the brightests among high-state CVs regardless of the subtypes \citep{war87CVabsmag, pat11CVdistance, ram17CVdistance}. On the other hand, the $g$-band absolute magnitude in quiescence is 12.1--12.5 mag, which is within the range of ordinary low mass-ratio CVs (e.g., \cite{iso19asassn14dx, tam20j2104}).

\subsubsection{Irradiated secondary case}
\label{sec6:irrparam}

\begin{figure}[tbp]
 \begin{center}
  \includegraphics[width=\linewidth]{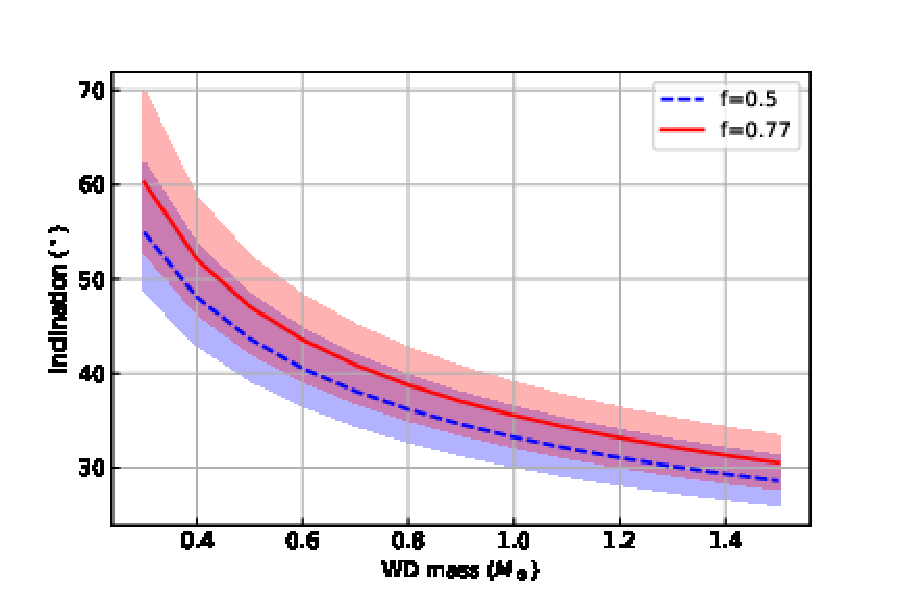}
 \end{center}
 \caption{
 Relation between the WD mass and the inclination under the irradiated secondary case.
 The red and blue lines represent the cases in which the k-correction factor $f_{\rm cor}$ is 0.5 and 0.77, respectively.
 The colored areas correspond to the one sigma error of the radial velocity amplitude of He \textsc{ii}~4686.
}
 \label{fig:rvtoincl}
\end{figure}

One possible interpretation of the observed S-wave component is an irradiated secondary star. We note that the distance, WD mass, and WD radius estimated in \citet{kat22WZSgecandle, kim23j0302} following the TTI case cannot be accepted in this case. This is because the expected parameter spaces in the irradiated secondary case do not necessarily align with those applied in the above papers. Since the observed radial velocity amplitude $K_{2,o}$ is not equal to the true radial velocity amplitude of the secondary star $K_2$ due to the unevenly irradiated surface of the secondary star, one needs the so-called k-correction \citep{wad88zcha}. We here adopted the observed radial velocity of He~\textsc{ii} 4686, as it has much smaller uncertainty than that of H$\beta$, and the k-correction factor $f_{\rm cor}=0.5$ and $0.77$ representing the optically thick and thin line, respectively \citep{par10nnser, par12irradiatedmdwarf}. The relation between $M_{\rm  WD}$ and $i$ is given under the Kepler rotation as equation \ref{eq:secrv}, using the gravitational constant $G$ \citep{neu23bwscl}.

\begin{equation}
    \label{eq:secrv}
    K_2 = \frac{K_{2,o}}{1 - 0.462 q^{1/3} \left(1 + q \right)^{2/3} f} = 
        \left( \frac{2 \pi G M_{\rm WD} \sin^3{i}}{P_{\rm orb} \left( 1+q \right)^2}\right)^{1/3}
\end{equation}

\noindent
Figure \ref{fig:rvtoincl} presents the possible range of $M_{\rm WD}$ and $i$ to interpret this S-wave as the irradiated secondary. Considering the relatively normal WD mass (0.6--1.0 M$_\odot$; \cite{pal22WDinCVs}), the inclination should be below 50$^\circ$, lower than the TTI case. The phase zero epoch equals the derived $T_0$ in section \ref{sec:5atmax}.

This irradiated secondary case, however, encounters some criticism. First, the binary parameters in this case are reliable only if the observed radial velocities purely represent that of the irradiated secondary. Any contamination from other emission-line sources easily gives the wrong values. Well-studied WZ Sge-type DNe in outburst maximum show compact He \textsc{ii} 4686-emitting structures, namely a structured disk \citep{bab02wzsgeletter, kuu02wzsge} and disk winds \citep{tam22v455andspec, tov22v455andspec}, raising a question on the reliability of the derived binary parameters in this case.
Second, although WZ Sge during the outburst decline ($\approx$ 20 d after the outburst maximum) showed the irradiated secondary in its Doppler map, it was detected only in H$\alpha$ \citep{ste01wzsgesecondary}. \citet{neu06bzuma} claim the detection of an irradiated secondary component in BZ UMa at the outburst maximum but only in Balmer and  He~\textsc{i}~lines. \footnote{\citet{neu06bzuma} do not discuss the inclination, WD mass, or mass ratio of BZ UMa utilized to plot the velocity of the secondary star on their Doppler map, which raises another question for their interpretation of the Doppler maps.} Other detections of the irradiated secondary component in the time-resolved spectra of high-state CVs are limited only in the low-excitation Balmer and He \textsc{i}~lines in our best knowledge (e.g., \cite{mar90ippeg, ste01wzsgesecondary, har05v3885sgr}). Hence, it is unclear if irradiation on the secondary star in high-state CVs is strong enough to predominantly produce a high-excitation line such as He \textsc{ii} 4686.
\footnote{It is worth noting that a radial velocity of the irradiated secondary has been detected in He \textsc{ii} 4686 and Bowen blend in cases of high-state low-mass X-ray binaries \citep{ste02scox1donor, cor08lmxbdonor, cor13lmxbdonor}, which usually show much stronger irradiating UV/X-ray emissions than high-state CVs do. Even in these cases, most systems have compact He \textsc{ii} 4686-emitting structure(s) other than the irradiated secondary, preventing {\em determination} of the binary parameters solely from the radial velocity of He \textsc{ii} 4686.}
Third, according to the TTI model and its expectation, the estimated inclination may be too low to produce the observed amplitude of early superhumps in MASTER J0302.
The most promising origin of early superhumps is the vertically-extended spiral-arm structure in disk excited by the 2:1 tidal resonance \citep{lin79lowqdisk, osa02wzsgehump, uem12ESHrecon}, and the variation is caused due to the orbital rotation of this structured disk. This means that low-inclination systems do not show detectable amplitude of early superhumps; \citet{kat22WZSgecandle} discussed that an amplitude of early superhumps becomes below 0.01 mag for inclinations lower than 40$^\circ$. If the inclination in this irradiated secondary case is correct, the origin of early superhumps in MASTER J0302 must be quite different from that in other WZ Sge-type DNe, in terms of its amplitudes and maxima phases.
Therefore, {\em without} knowing the inclination, WD mass, and orbital ephemeris, this irradiated secondary case may give extensively wrong binary parameters.

\section{Discussion}
\label{sec:7}

\subsection{MASTER OT J030227.28+191754.5 as a WZ Sge-type dwarf nova}
\label{sec:7asDN}

As described in section \ref{sec:3}, the 2021-2022 outburst of MASTER J0302 reached an amplitude of 10.2 mag and a duration of 60 d, much more energetic than other DNe regardless of subtypes (e.g., \cite{kaw21DNCNinASASSN}).  Indeed, \citet{ste21J0302specatel15069, tag21J0302specatel15072} suspected it could be a He nova eruption based on its large outburst amplitude and earliest optical spectra. We have detected double-peaked optical emission lines later (section \ref{sec:5}), and early and ordinary superhumps  (section \ref{sec:4SH}). The color temperature during the outburst of MASTER J0302 exceeds 10000 K which is consistent with other DN outbursts (\cite{shu21aylac} and references therein),  while is much hotter than $\sim$ 8000 K expected around the peak of classical nova eruptions \citep{bec95novashell, war95book}. These points establish that the origin of the MASTER J0302 outburst is thermal--tidal instability in an accretion disk.

\begin{figure}[tbp]
 \begin{center}
  \includegraphics[width=\linewidth]{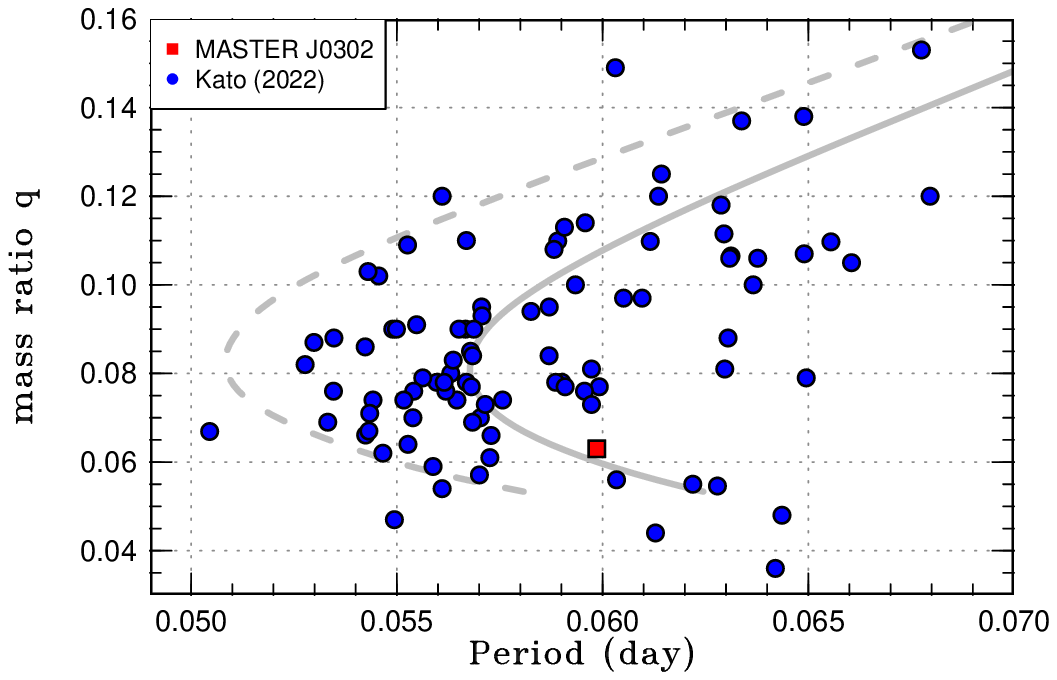}
 \end{center}
 \caption{The orbital period v.s. the mass ratio of WZ Sge-type DNe.
 The red square represents MASTER J0302, and the blue circles show other WZ Sge-type DNe from \citet{kat22updatedSHAmethod}.
 The dashed and solid lines show the theoretical and empirical evolutionary tracks, respectively, from \citet{kni11CVdonor}.}
 \label{fig:q_p}
\end{figure}

Despite its extreme outburst energetics, its superhump periods and mass ratio are in the range of ordinary low mass-ratio CVs [\citet{kat15wzsge, kat22updatedSHAmethod} and references therein]. Figure \ref{fig:q_p} shows the relation between the orbital period and mass ratio of MASTER J0302 (red squares) and CVs listed in \citet{kat22updatedSHAmethod} (blue circles). The dashed and solid lines correspond to the theoretical and empirical evolutionary paths from \citet{kni11CVdonor}.  In this orbital period-mass ratio plane, MASTER J0302 is located along the standard evolutionary track of CVs and seems to be a period bouncer (i.e., hosting a degenerate secondary). However, it is worth noting that the bouncing {\em mass ratio} should depend on the WD mass, as the critical secondary mass for having a degenerate core ($\approx 0.06 M_\odot$; \cite{kni11CVdonor}) would be somewhat a constant.
Applying the WD mass ($\gtrsim 1.15 M_\odot$) derived by \citet{kim23j0302} yields the bouncing mass ratio $\lesssim$ 0.052, smaller than the estimated one for MASTER J0302. Hence the secondary star cannot have a degenerate core in this WD mass range.
It is an open question how such a low mass-ratio CV with an ONe (and massive; $\gtrsim 1.15$M$_\odot$) WD is formed through the evolution of CVs.

\subsection{Outburst physics of MASTER OT J030227.28+191754.5}
\label{sec:7inWZSge}

In this subsection, we summarize possible interpretations of the MASTER J0302 outburst. As revealed in section \ref{sec:4SH}, the superhump behaviors in MASTER J0302 are reasonably consistent with ordinary WZ Sge-type DNe. Therefore, it is natural to figure out what aspect(s) in the widely accepted TTI model explains the outburst energetics of MASTER J0302. Our following discussion is based on the scaling and extensions of the theoretically and empirically obtained relations in DN outbursts.  Detailed simulations for reproducing the observed light curve and their characteristics are beyond the scope of this paper.

\subsubsection{10.2 mag outburst amplitude}
\label{sec:72amplitude}

Typical outburst amplitudes of WZ Sge-type DNe are 6--8 mag \citep{kat15wzsge}. Their absolute magnitudes at outburst maximum and in quiescence cluster around 3.0--6.0 and 11.0--13.0 mag, respectively \citep{tam20j2104}.  To explain the outburst amplitude reaching 10.2 mag in MASTER J0302, either brighter outburst maximum and/or fainter quiescence than other WZ Sge-type DNe is required.

\begin{figure}[tbp]
 \begin{center}
  \includegraphics[width=\linewidth]{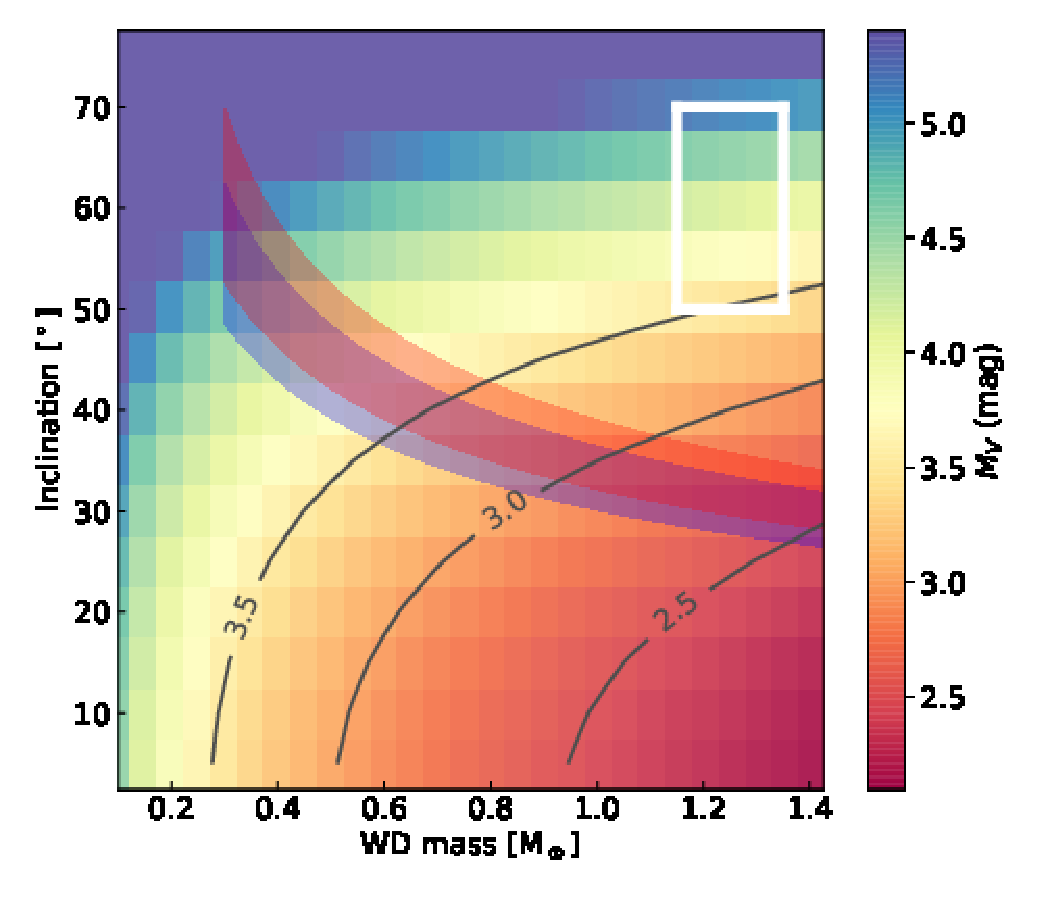}
 \end{center}
 \caption{The expected $V$-band absolute magnitude $M_V$ of the standard disk model around the WD with the $1\times10^{18}~{\rm g}~{\rm s}^{-1}$ accretion rate, on the WD mass--inclination plane.
 See the text for other parameters adopted for the calculation.
 The white box and the shaded area represent the estimated binary parameter range in the TTI case and the irradiated secondary case (red and blue for $f_{\rm cor} = $ 0.5 and 0.77, respectively).}
 \label{fig:absmag}
\end{figure}

To assess this, we first check the expected absolute magnitudes of DNe in outburst on the WD mass $M_{\rm WD}$ and inclination $i$ plane. Optical continuum of DNe at outburst maximum is reasonably approximated with the standard disk model \citep{sha73alphadisk} around the primary WD with an accretion rate $\dot{M}_{\rm acc}~\approx~5 \times 10^{17}$-- $1 \times 10^{18} ~{\rm g}~{\rm s}^{-1}$ (e.g., \cite{hor85zcha, bru96oycareclipsemapping}). The outburst model of WZ Sge by \citet{osa95wzsge} also gives a similar accretion rate at outburst maximum.  Following these, we calculated the expected $V$-band absolute magnitude $M_V$ at the accretion rate $\dot{M}_{\rm acc}~=~1\times10^{18}~{\rm g}~{\rm s}^{-1}$ using the standard disk model and the equations presented in Appendix \ref{sec:alphadisk}, with the fixed orbital period $P_{\rm orb}=0.05986$ d and mass ratio $q=0.063$. We assumed that the inner and outer disk radii are truncated at the WD radius \citep{lam17book} and the 2:1 resonance radius \citep{osa02wzsgehump}, respectively. We present this result in figure \ref{fig:absmag}. The typical WD mass ($0.8~{\rm M}_\odot$) and average inclination (1 radian $=57^\circ$) yields $M_V \approx 4.2$ mag, indeed consistent with the typical absolute magnitude of WZ Sge-type DNe at outburst maximum.

The white box in figure \ref{fig:absmag} represents the estimated binary parameter range in the TTI case. The expected $M_V$ in the TTI case is all fainter than 3.0 mag, which is in short to explain the observed absolute magnitude at outburst maximum ($M_V \simeq$ 2.4--2.2) and outburst amplitude. The only way to explain this with the standard disk model is to enhance the accretion rate. Given the binary parameters in the TTI case, the accretion rate explaining its luminosity at outburst maximum is $\dot{M}_{\rm acc} \approx 1 \times 10^{20}~{\rm g}~{\rm s}^{-1}$, $\sim$2 orders of magnitude higher than the typical ones.

The shaded area in figure \ref{fig:absmag} represents the estimated binary parameter range in the irradiated secondary case.  We repeat that there is no constraint on the distance and WD mass in this case. The absolute magnitude between 2.5 -- 3.0 mag is expected in the space where the WD mass is relatively massive ($\gtrsim 0.9$ M$_\odot$) and the inclination is low ($\lesssim 40^\circ$). Therefore, once the quiescence luminosity of MASTER J0302 is comparable to the fainter end ($\geq 12.5$ mag) of WZ Sge-type DNe, the 10.2-mag outburst amplitude can be explained simply with the standard accretion rate but a disk around a massive WD in low inclination.

The fainter quiescence in MASTER J0302 is unlikely based on the distance in the TTI case. This distance gives an absolute magnitude in quiescence as $M_{\rm g}$ = 12.5--12.1 mag, typical as a WZ Sge-type DN. Moreover, the de-reddened colors of MASTER J032 in GALEX and SDSS are compatible with those of WZ Sge-type DNe \citep{wil10newCVs, kat12DNSDSS}. The GALEX color $FUV-NUV=-0.3(5)$ align with ones from $\log{g}=8$ WDs with $\gtrsim 12000$ K \citep{ren20gaiawdbinary}. Since spectral energy density (SED) of WZ Sge-type DNe in quiescence is dominated by a primary WD (e.g., \cite{kni11CVdonor}), a lower temperature of the WD is essential to accomplish fainter quiescence magnitude. Thus, even in the irradiated secondary case, these points support that the quiescence brightness of MASTER J0302 is comparable to other WZ Sge-type DNe, rather than $\geq 2$ mag fainter than even their fainter ends.

\subsubsection{60-d long outburst duration}
\label{sec:72duration}

\begin{figure}[tbp]
 \begin{center}
  \includegraphics[width=\linewidth]{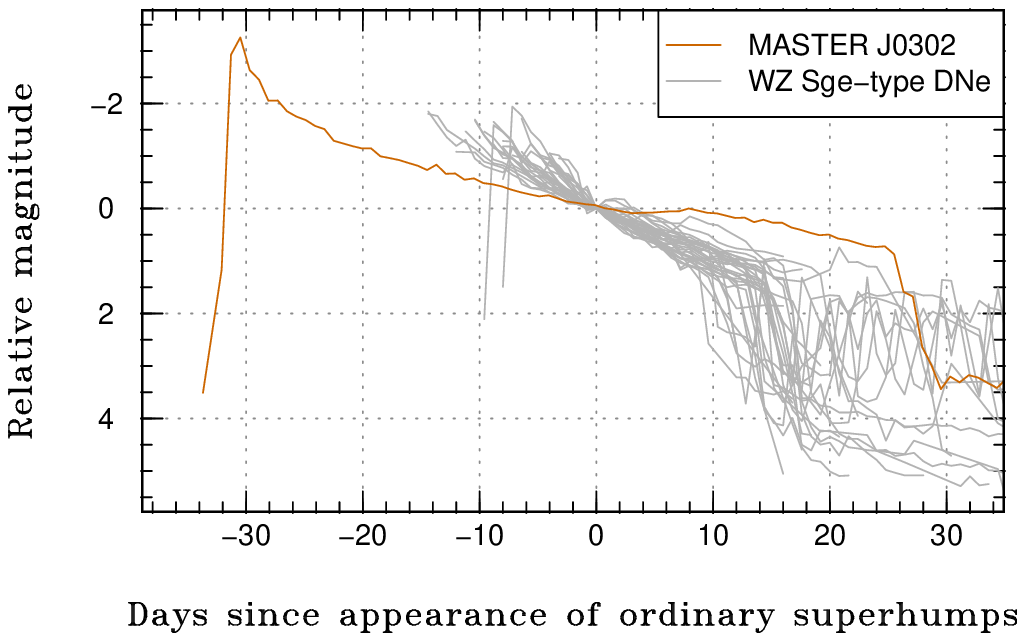}\\
 \end{center}
 \caption{
 The outburst light curves of MASTER J0302 (orange) and other WZ Sge-type DNe (gray) listed in table 1 of \citet{kat22WZSgecandle}.
 All the data are binned in 1 d.
 The light curve is shifted so that the magnitude and epoch at the appearance of ordinary superhumps, which is suggested to be a standard candle \citep{kat22WZSgecandle}, match each other.
 }
 \label{fig:compLC}
\end{figure}

Another uniqueness of MASTER J0302 is its long superoutburst duration. Figure \ref{fig:compLC} compares the outburst light curves of MASTER J0302 (orange) and other WZ Sge-type DNe (gray) listed in table 1 of \citet{kat22WZSgecandle}. The 60-d duration of the MASTER J0302 outburst is almost double -- triple ones of other WZ Sge-type DNe (20--30 d).

Since DN outbursts are powered by accretion, the duration of an outburst is largely determined by mass accreted during the outburst \citep{osa89suuma}. A massive disk in quiescence of MASTER J0302 is favored from observations: Since the early superhump phase is interpreted as the viscous depletion phase \citep{osa95wzsge, can01wzsge},  the intrinsically ($\geq$ 30 d) and relatively (exceeding a half of the outburst duration) long early superhump phase of MASTER J0302 imply a larger stored mass in the disk at the onset of the outburst than that of other WZ Sge-type DNe.

Maximum disk mass $M_{\rm disk,~max}$ that does not trigger an outburst by thermal instability is given in equation 3.26a of \citet{war95book}  using the viscosity of the disk in quiescence $\alpha_{\rm cool}$, the WD mass $M_{\rm WD}$, and outer disk radius in quiescence $r_{\rm cool}$ (equation \ref{eq:diskmass}).

\begin{equation}
    \label{eq:diskmass}
    M_{\rm disk,~max} \propto {\alpha_{\rm cool}}^{-0.86} ~ {M_{\rm WD}}^{-0.35} ~ {r_{\rm cool}}^3
\end{equation}

\noindent
Since the orbital period and mass ratio are obtained by our observations, the disk radius ${r_{\rm cool}}$ can be rewritten as ${r_{\rm cool}}^3 \propto M_{\rm WD}$ by considering a Keplerian disk. Therefore, the dependence of maximum disk mass on the primary WD mass is $M_{\rm disk,~max} \propto {\alpha_{\rm cool}}^{-0.86}{M_{\rm WD}}^{0.65}$. Even though a more massive WD will result in more mass in a disk, the quiescence viscosity has a larger dependence on the disk mass, whose exact value and origin in WZ Sge-type DNe are still not well understood.

\subsubsection{Possible origin of the MASTER OT J030227.28+191754.5 superoutburst}
\label{sec:7scenario}

\begin{figure}[tbp]
 \begin{center}
  \includegraphics[width=\linewidth]{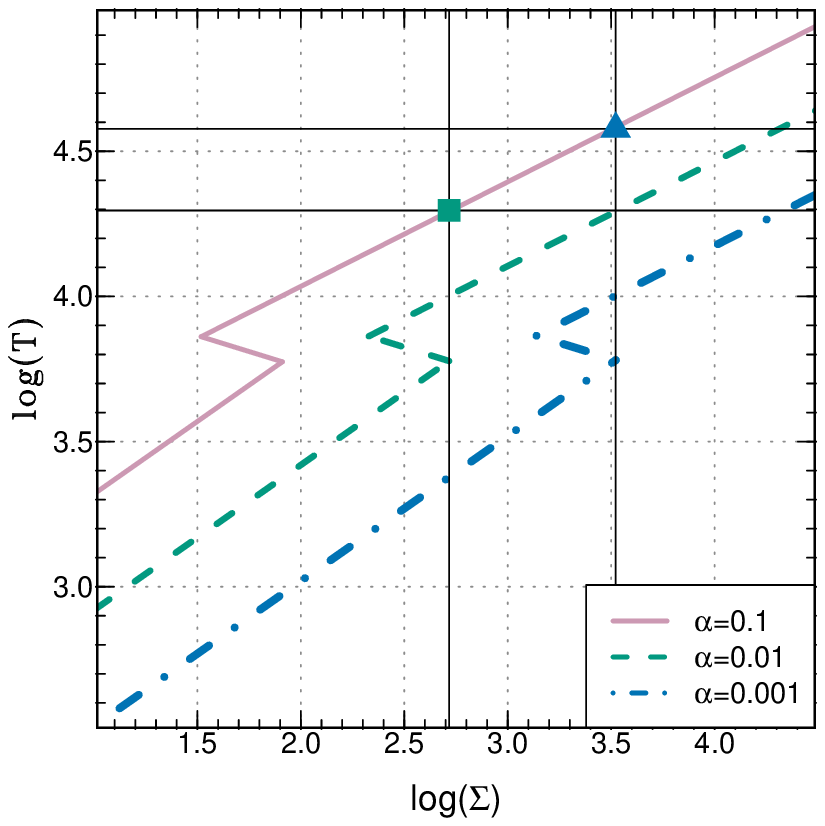}\\
 \end{center}
 \caption{
 Schematic image of the thermal equilibrium curves at a specific disk radius $r$ with different viscosity $\alpha$, adopted in the thermal instability model.
 X and Y axes correspond to the disk surface density $\Sigma$ and temperature $T$.
 The pink solid, green dashed, and blue dot-dashed lines show the equilibrium curves applying the viscosity $\alpha = $ 0.1, 0.01, and 0.001, respectively.
 See text for other adopted binary parameters.
 The green square and blue triangle show the maximum surface density and temperature in the limit cycles between the high-state viscosity $\alpha_{\rm hot} = 0.1$, and the low-state viscosity $\alpha_{\rm cool} = 0.01$ and $0.001$, respectively.
 }
 \label{fig:DIscheme}
\end{figure}

As we discussed in this subsection, the more massive a primary WD is, the longer and brighter the DN outburst is.  In the case of MASTER J0302, \citet{kim23j0302} discussed that this system would host the massive (1.15--1.34 M$_\odot$) primary WD. However, in the TTI case, our previous discussion suggests that this massive WD cannot solely explain the outburst energetics. Indeed another well-studied WZ Sge-type DN hosting a relatively massive WD is 1RXS J023238.8$-$371812, whose WD mass is estimated as 1.15$\pm{0.04}$ M$_\odot$ \citep{pal22WDinCVs}. This underwent the superoutburst in 2007, with the outburst amplitude and duration of just $8.2$ mag and $\sim 15$ d, respectively \citep{Pdot}.

One plausible parameter that can achieve both aspects is quiescence viscosity even lower than other WZ Sge-type DNe. Figure \ref{fig:DIscheme} illustrates the schematic image of the thermal equilibrium curves in the disk surface density $\Sigma$ -- temperature $T$ plane. Formulae plotting these curves are taken from equation 5.19--5.21 of \citet{kat08book}, with the WD mass $M_{\rm WD} = 1.25~{\rm M}_\odot$ and at a specific disk radius $r = 10^{10}~{\rm cm}$ (corresponding to {$\approx$ 34 $\times~R_{\rm WD}$}). The limit cycle explaining DN outbursts applies two different viscosity $\alpha_{\rm hot}$ and $\alpha_{\rm cool}$ for outburst and quiescence, respectively. The outburst variety between DN subtypes is explained by the different low-state viscosity \citep{min85DNDI, osa95wzsge}.
Considering this limit cycle, the smaller quiescence viscosity, from the dashed ($\alpha_{\rm cool} = 0.01$) to dot-dashed ($\alpha_{\rm cool} = 0.001$) lines, leads to larger critical surface density to trigger an outburst, and hotter disk temperature at outburst maximum corresponding to this surface density, from the green square to the blue triangle in figure \ref{fig:DIscheme}. This hotter disk temperature results from a higher accretion rate, which explains the bright outburst maximum and large outburst amplitude of MASTER J0302. It is worth noting that this discussion would still be qualitatively eligible in the parameter spaces other than the TTI case as long as the expected absolute magnitude in figure \ref{fig:absmag} is fainter than $\simeq 3$ mag. Therefore, while future dedicated simulations will be preferred to confirm what degree of lowering quiescence viscosity can realize its outburst amplitude and duration, the quiescence viscosity even lower than ordinary WZ Sge-type DNe gives a natural explanation both for the bright outburst maximum and long outburst duration of MASTER J0302 within the framework of the current TTI model.

On the other hand, such an enhanced accretion rate may not necessarily be required in a part of the irradiated secondary case. Inclination low enough ($i \leq 40^\circ$) and WD mass high enough ($\leq 0.9~{\rm M}_\odot$) lead brighter {\em observed} absolute magnitudes possibly explaining outburst amplitude. Therefore, in this case, only the disk mass stored before the outburst onset would matter, which can be explained solely by the result of a large disk around a massive WD. Nevertheless, it should be finally acknowledged that, in the irradiated secondary case, early superhump amplitude and phasing of its maxima in MASTER J0302 do not agree with those in other WZ Sge-type DNe. Understanding this while keeping consistency with other WZ Sge-type DNe would require severe updates in the TTI model.

\subsection{Doppler tomography and its interpretation}
\label{sec:6Dopmap}

\begin{figure*}[tbp]
 \begin{center}
  \includegraphics[width=0.9\linewidth]{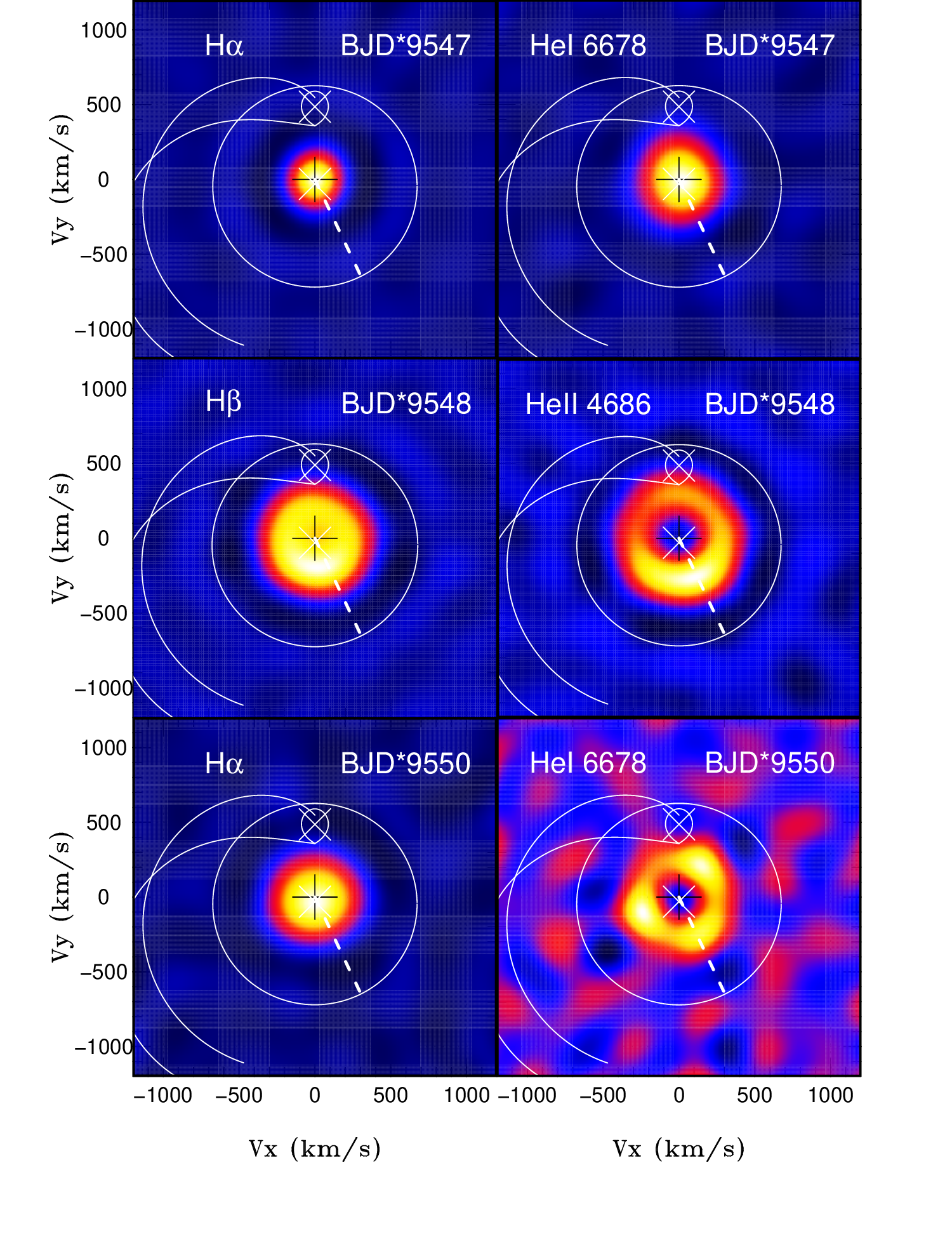}
 \end{center}
 \caption{Doppler tomography of MASTER J0302 during its outburst; 
 upper left:  H$\alpha$ on BJD 2459547.2, 
 upper right:  He~\textsc{i} 6678 on BJD 2459547.2, 
 middle left:  H$\beta$ on BJD 2459548.2,
 middle right:  He~\textsc{ii} 4686 on BJD 22459548.2,
 lower left:  H$\alpha$ on BJD 2459550.1, and 
 lower right:  He~\textsc{i} on BJD 2459550.1, 
 The secondary star, the mass stream, and the corresponding Kepler velocity of the tidal truncation radius are presented with the solid lines using the binary parameters obtained in the TTI case and assuming $M_{\rm WD} = 1.25$ M$_{\odot}$.
 The dashed lines represent the orbital phase at which the orbital phase is zero in the irradiated secondary case. 
}
 \label{fig:dopmap}
\end{figure*}

Doppler tomography is a classical method to analyze the system structure in compact binaries from phase-resolved emission lines  \citep{DopplerTomography}. We used the Doppler tomography code developed by \citet{uem15DTTVM} and applied it to the phase-resolved spectroscopic data of H$\alpha$, H$\beta$, He~\textsc{i} 6678, and He~\textsc{ii} 4686 obtained on BJD 2459547.2, 2459548.2, and 2459550.1 (figure \ref{fig:phasespec}). We utilized the orbital ephemeris of equation \ref{eq:ephemeris} to create the Doppler map contour.

The obtained Doppler maps are presented in figure \ref{fig:dopmap}. The observation epoch and analyzed line are presented at the top of each panel. The Roche lobe of the secondary star and the mass stream from the Lagrangian point $L_1$ are presented based on the binary parameters obtained in the TTI case and assuming $M_{\rm WD} = 1.25$ M$_{\odot}$. The solid circle (670 km s$^{-1}$) corresponds to the Kepler velocity at the tidal truncation radius \citep{pac77ADmodel} obtained from these binary parameters. The white dashed lines represent the orbital phase shift from the TTI case at which the orbital phase is zero in the irradiated secondary case.

The Doppler maps of H$\alpha$ and He~\textsc{i} 6678 on BJD 2459547, and H$\alpha$ on BJD 2459550 show a compact structure centered at the primary WD and slower than 300 km s$^{-1}$. Meanwhile, those of H$\beta$ and He~\textsc{ii} 4686 on BJD 2459548 and He~\textsc{i} 6678 on BJD 2459547 shows a ring structure with the typical velocity of $\sim 300$ km s$^{-1}$. The He~\textsc{ii} 4686 and H$\beta$ Doppler maps show the arc-shaped asymmetric structure around $(V_X, V_Y) \simeq (100, -250)$ km s$^{-1}$, which was seen as a S-wave structure in the time-resolved spectra. All these velocities are slower than the expected rotational velocity at the tidal truncation radius in the TTI case. Even in the irradiated secondary case, any emission lines originating from a Keplerian disk should have faster rotation velocity (by a factor of $\geq$1.3 for $q=0.063$) than the (un-irradiated) secondary star. These typical velocities (peak separation of $\leq 400$ km s$^{-1}$) are comparable to or even smaller than that of the S-wave structure (full amplitude of $\simeq 500$ km s$^{-1}$), indicating that, in either case, the main line-forming source of these emission lines should not be a rotating disk, at least for the symmetric component.

These compact structure patterns on Doppler maps, together with the strong high-excitation emission lines, share similarities to the outburst spectra of a WZ Sge-type DN V455 And (\cite{tam22v455andspec}, later confirmed by \cite{tov22v455andspec}), and those of so-called SW Sex-type CVs \citep{tho91bhlyn, sch15swsex}. Among some proposed interpretations for this feature, a magnetic curtain around the primary WD \citep{wil89CVeclipse, cas96v795her, hoa03dwuma} is unlikely due to the non-detection of the WD spin period in neither X-ray and optical. A structured hotspot (e.g., \cite{dhi13swsexenigma}) is also implausible because the observed compact structures in H$\alpha$ do not show any radial velocity. Therefore, the most probable interpretation is disk winds, as for V455 And (\cite{tam24dnwind} and references therein). MASTER J0302 is thus classified as the rare DNe showing clear evidence of disk winds in its optical spectra in outburst.

Disk winds would also explain the observed emission line features. According to the numerical simulations by \citet{mat15CVdiskwind}, Balmer and He~\textsc{ii} lines are the most contributed by disk winds. This trend is seen in the spectra of MASTER J0302 and V455 And compared to other WZ Sge-type DNe (\cite{tam21seimeiCVspec} and references therein). Especially in the TTI case, the accretion rate of MASTER J0302 should be orders higher than those of other WZ Sge-type DNe (section \ref{sec:7}).
\citet{tam24dnwind} discuss that such a high accretion rate can be a key for emerging strong disk wind signatures in optical spectra.
The profile difference between line species, like between Balmer and He \textsc{ii}~lines at the same epoch, can be interpreted as the different line opacities \citep{rib08CVflickering, mat15CVdiskwind, tov22v455andspec}. Simulations in \citet{tam24dnwind} also showed that the decreasing accretion and mass-loss rates can naturally explain the transition of H$\alpha$ from single-peaked to double-peaked profile along outburst decline.

\section{Summary}
\label{sec:8}

We report our intensive optical-IR observation campaign during an outburst of MASTER J0302 in 2021-2022. The outburst amplitude and duration reached 10.2 mag and 60 d, respectively. Our detections of double-peaked emission lines and early superhumps established that MASTER J0302 is a disk-powered but enormously energetic WZ Sge-type DN.

The periods of early and growing ordinary superhumps were obtained as 0.05986(1) and 0.06131(2) d, respectively. Using these two superhump periods, we estimate its mass ratio to be 0.063 (1), below typical ones at the period minimum. Its superhump properties are all consistent with expectations as a low-mass-ratio DN. In terms of binary evolution, the early superhump period and mass ratio of MASTER J0302 follow the standard evolutionary path.  However, its secondary star may not be degenerated if the WD is massive ($\geq 1.15~{\rm M}_\odot$) as inferred from X-ray observations.

We derived the binary parameters of MASTER J0302 in two cases; (1) MASTER J0302 follows the same superhump behaviour as other WZ Sge-type DNe (the TTI case), and (2) the origin of the S-wave structure in the time-resolved spectra is the irradiated secondary (the irradiated secondary case). According to the binary parameters in the TTI case, the highly enhanced accretion rate ($\approx 10^{20}~{\rm g}~{\rm s}^{-1}$) is needed to explain the large outburst amplitude. Even in the irradiated secondary case, the enhancement of the accretion rate would be favored unless the WD is relatively massive ($\geq 0.9~{\rm M}_\odot$) and the inclination is low ($\leq 40^\circ$). At such a low inclination, the origin of early superhump can be greatly different from other systems. Along with this, the long outburst duration needs a large disk mass stored in quiescence. In the thermal-tidal instability model, one possible factor that can explain both these aspects is quiescence viscosity even lower than other WZ Sge-type DNe. 

The spectra of MASTER J0302 around the outburst maximum showed strong emission lines of Balmer, He~\textsc{i}, He~\textsc{ii}, and other high-excitation series. Typical peak separation and FWHM of these line profiles are considerably narrower than those expected from a Keplerian accretion disk in either case, needing the other line-forming source. The similarity with V455 And and SW Sex-type CVs suggests that disk winds are the most plausible origin for these emission lines.

\begin{ack}

This work was financially supported by the Japan Society for the Promotion of Science Grants-in-Aid for Scientific Research (KAKENHI) Grant Numbers 21J22351 (Y.T.), 21K03616 (D.N. and T.K.), JP20K22374 (M.K.), JP21K13970 (M.K.), 22K03676 (M.Y.), and 19K03930 (K.M.).  
M. K. acknowledges support from the Special Postdoctoral Researchers Program at RIKEN.  
This research is partially supported by the Optical and Infrared Synergetic Telescopes for Education and Research (OISTER) program funded by the MEXT of Japan.
JSPS Grant-in-Aid for Scientific Research on Innovative Areas (17H06362), 
and the joint research program of the Institute for Cosmic Ray Research (ICRR), 
MEXT/JSPS KAKENHI Grant Numbers 15H02063, JP18H05439, P17H04574, 18H05442, and 22000005, 
and JST CREST Grant Number JPMJCR1761. 
The authors thank the TriCCS developer team (which has been supported by the JSPS KAKENHI grant Nos. JP18H05223, JP20H00174, and JP20H04736, and by NAOJ Joint Development Research).
This work was supported by the Slovak Research and Development Agency under the contract No. APVV-20-0148

Lasair is supported by the UKRI Science and Technology Facilities Council and is a collaboration between the University of Edinburgh (grant ST/N002512/1) and Queen’s University Belfast (grant ST/N002520/1) within the LSST:UK Science Consortium. ZTF is supported by National Science Foundation grant AST-1440341 and a collaboration including Caltech, IPAC, the Weizmann Institute for Science, the Oskar Klein Center at Stockholm University, the University of Maryland, the University of Washington, Deutsches Elektronen-Synchrotron and Humboldt University, Los Alamos National Laboratories, the TANGO Consortium of Taiwan, the University of Wisconsin at Milwaukee and Lawrence Berkeley National Laboratories. Operations are conducted by COO, IPAC, and UW. This research has made use of ``Aladin sky atlas'' developed at CDS, Strasbourg Observatory, France \cite{Aladin2000, Aladinlite}.
We acknowledge ESA Gaia, DPAC and the Photometric Science Alerts Team (http://gsaweb.ast.cam.ac.uk/alerts).

\end{ack}

\section*{Supporting Information}
The following supporting information is available in the online version of this article: Tables E1--E6 and Figure E1--E3.

\appendix

\setcounter{figure}{0}
\setcounter{table}{0}
\renewcommand{\thefigure}{A\arabic{figure}}
\renewcommand{\thetable}{A\arabic{table}}

\section{Orbital ephemeris based on early superhump profiles}
\label{sec:eshephemeris}

In this appendix section, we describe our method to determine the orbital ephemeris of MASTER J0302 based on its early superhump profile. In WZ Sge-type DNe, the early superhump period has been adopted as the orbital period \citep{kat15wzsge}. The profiles of early superhumps are almost inherent to the systems \citep{kat02wzsgeESH, kat15wzsge}, as early superhump maxima are determined by the phase of double-arm structures excited by the 2:1 tidal resonance in the TTI model. This is also observationally supported \citep{uem12ESHrecon, nak13j0120, tam22PNVJ0044+41}. Therefore, the orbital ephemeris of a WZ Sge-type DN can be determined by comparing the out-of-eclipse profile of early superhumps with those of the systems whose orbital ephemeris (i.e., mid-eclipse time of the WD and disk eclipse) has been already known from eclipse light curves and/or radial velocities. In this paper, we applied WZ Sge \citep{pat98wzsge, ste01wzsgesecondary, ish02wzsgeletter}, BW Scl \citep{Pdot4, neu23bwscl}, MASTER OT J005740.99+443101.5 (MASTER J0057; \citet{Pdot6}), and OV Boo \citep{pat08j1507, kat22WZSgecandle} as the template samples (table \ref{tab:a1}).
\citet{pat08j1507} measured the mid-eclipse timing of OV Boo based on the WD eclipse. The orbital ephemeris of WZ Sge \citep{ste01wzsgesecondary}  and BW Scl \citep{neu23bwscl} are obtained from the radial velocity studies of the irradiated secondary, whose phase zero timing should correspond to the mid-eclipse time of the WD.  \citet{Pdot6} determined the orbital ephemeris of MASTER J0057 based on the disk eclipses during the ordinary superhump and post-outburst phase. In these epochs, the most dominant light source is the symmetric accretion disk, and hence the averaged mid-eclipse timing should correspond to the WD mid-eclipse. It is worth noting that we did not include the systems as the template whose ephemeris is determined based on the grazing eclipse. This is because, as discussed in \citet{spr98wzsge, ste01wzsgesecondary}, the mid-time of the grazing eclipse in WZ Sge is systematically shifted by $\sim$0.04 orbital phase from the true disk eclipse time due to the contribution from a hot spot.

\begin{table*}
\caption{Applied orbital ephemeris of template objects}
\centering
\label{tab:a1}
\begin{tabular}{ccccc}
  \hline              
       &  $T_0$ [BJD - 2400000] & $P_{\rm orb}$ [d] & Method & Reference \\
    \hline
    WZ Sge   &  37547.72891(4) & 0.05668784689(4) & radial velocity & \citet{ste01wzsgesecondary} \\
    BW Scl   &  50032.13628(11) & 0.0543239136(24) & radial velocity & \citet{neu23bwscl} \\
    MASTER J0057   &   56617.36772(4) & 0.0561904(3) & disk eclipse & \citet{Pdot6} \\
    OV Boo   &   53498.892253(9) & 0.0462583411(7) & WD eclipse & \citet{pat08j1507} \\
  \hline
\end{tabular}
\end{table*}

For the quantitative estimation of the phase zero epoch of MASTER J0302, we applied the Bayesian technique. We denote  the observed early superhump magnitude of MASTER J0302 as $\boldsymbol{x}=(x_1, \ldots, x_n)$ with the standard deviation error of $\boldsymbol{\delta}=(\delta_1, \ldots, \delta_n)$ at the observed epoch of $\boldsymbol{t_x}=(t_{x_1}, \ldots, t_{x_n})$. To avoid the contamination of eclipses in matching the early superhump profiles, we excluded the orbital phase $-$0.15 -- 0.1 from our calculation. As the first step of the calculation, we created the model function $f_{\rm ESH} (t')$ of each template object by interpolating the observed early superhump magnitude as the function of the orbital phase $t'$ through the linear spline interpolation. Our Bayesian model assumes that the observed magnitudes of MASTER J0302 are given from the model function parameterized with the amplitude factor $a$ and the orbital phase difference $\Delta t$ from our assumed mid-eclipse time $T_0$ = BJD 2459550.1340 of MASTER J0302.

\begin{figure}[tbp]
 \begin{center}
  \includegraphics[width=\linewidth]{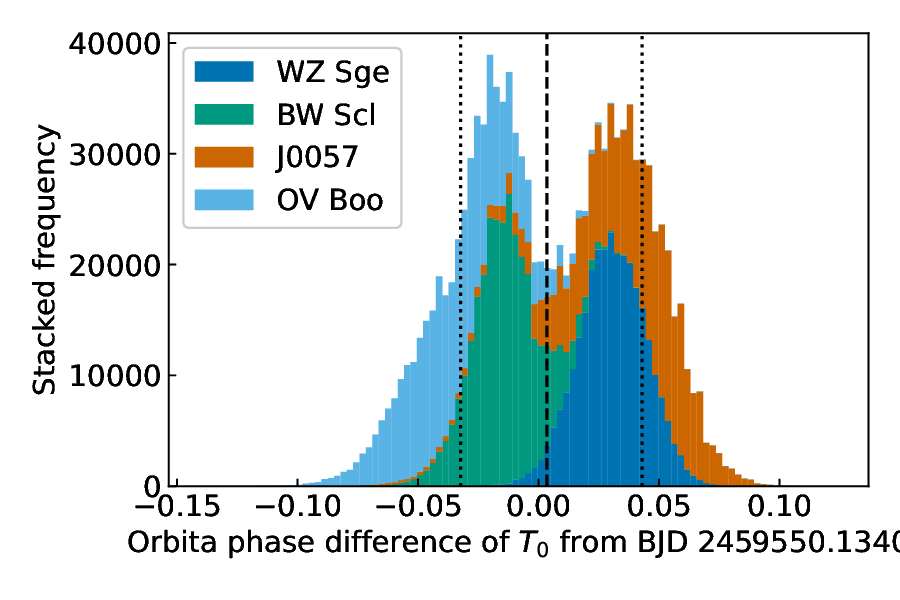}
 \end{center}
 \caption{
 The stacked distribution of the orbital phase difference of $T_0$ from our assumed $T_0$ epoch BJD 2459550.1340.
 The different colors show the different template objects; WZ Sge (blue), BW Scl (green), MASTER J0057 (brown), and OV Boo (cyan).
 The dashed and dotted lines show the median value and 64 percent quantile-based error estimated from our template samples.
 }
 \label{fig:esh-CCC}
\end{figure}

Therefore, by assuming that the observed magnitudes are generated from the Gaussian distributions $Norm$ centered at the latent magnitudes with standard deviations of the measurement errors, the observed magnitude of MASTER J0302 $\boldsymbol{x}$ can be written;

\begin{equation}
    \boldsymbol{x}_i \sim Norm \left( a \times f_{\rm ESH} \left( \frac{\boldsymbol{t_{xi}} - T_0}{P_{\rm orb}} + \Delta t \right), \boldsymbol{\delta}_i \right)
\end{equation}

We set the prior distribution as $\Delta t \sim {\rm Uniform}(-0.5, 0.5)$ and $a \sim {\rm Uniform}(0, 1)$ (i.e., the early superhump amplitude of the template objects are larger than that of MASTER J0302). We performed the Markov Chain Monte Carlo (MCMC) calculation for 1000000 iterations and discarded the initial 700000 iterations as burn-in, to determine the best parameters which minimize the difference between the model and the observation. Figure \ref{fig:esh-CCC} shows the result of the stacked $\Delta t$ distribution of our template objects from our MCMC calculation. We determined the median $\Delta t$ and the mid-eclipse time of MASTER J0302 from the $\Delta t$ distribution of four template objects. The error of $T_0$ was estimated as the one sigma error from the 68 percent quantile-based interval.  The obtained ephemeris is equation \ref{eq:ephemeris};

\begin{equation}
    T_0 = {\rm BJD~}   2459548.0391^{+0.0024}_{-0.0021} + 0.05986 \pm 0.00001 \times E
    \label{eq:ephemeris}
\end{equation}

One might see the $\Delta t$ distribution as bimodal, however, this is likely due to our limited number of template samples. We note that, in passing, this method may give a half-phase shift due to the double-peaked profile of early superhumps.

\section{The standard disk model in high-state CVs}
\label{sec:alphadisk}

{The standard disk model is generally accepted to explain the overall SED from UV to optical in high-state CVs (i.e., novalike variables and DNe in outburst).} In the standard disk model \citep{sha73alphadisk}, the effective temperature $T_{\rm eff}(r)$ of the accretion disk at a radius $r$ is written as equation \ref{eq:disktemp},

\begin{equation}
    \label{eq:disktemp}
    T_{\rm eff}(r) = \left[ \frac{3 G M_{\rm WD} \dot{M}_{\rm acc}}{8 \pi \sigma r^3} \left( 1 - \sqrt{\frac{r_{\rm in}}{r}}\right)\right]^{1/4},
\end{equation} 

\noindent
where  $G$, $\sigma$, $M_{\rm WD}$, $\dot{M}_{\rm acc}$, $r_{\rm in}$ are the gravitational constant, the Stefan–Boltzmann constant, the primary mass, the accretion rate in the disk, and the inner disk radius, respectively. In our calculation, we assumed that the inner disk reaches the WD surface and therefore $r_{\rm in} = R_{\rm WD}$. The disk locally radiates as a blackbody with effective temperature $T_{\rm eff}(r)$, and hence blackbody radiation $B_\nu(r)$ at a radius $r$ and observed flux $S_\nu$ at a frequency $\nu$ is given as equation \ref{eq:blackbody} and \ref{eq:standardisk} using the Planck constant $h$, the speed of light $c$, the Boltzmann constant $k$, the inclination of the system $i$, the distance of the system $D$, and the outer disk radius $r_{\rm out}$. In the case of WZ Sge-type DNe, the outer disk radius is considered to be truncated at the 2:1 resonance radius $r_{2:1}$, and we applied $r_{\rm out} = r_{2:1}$.

\begin{equation}
    \label{eq:blackbody}
    B_\nu(r) = \frac{2h}{c^2} \frac{\nu^3}{\exp{[h\nu/k T_{\rm eff}(r)]}-1} \\
\end{equation} 

\begin{equation}
    \label{eq:standardisk}
    S_\nu = \frac{\cos{i}}{D^2} \int^{r_{2:1}}_{R_{\rm WD}} B_\nu(r) 2 \pi r dr
\end{equation} 

\noindent
We also take into account the limb darkening effect $\Delta M$ of the optically thick disk depending on the inclination $i$ as in equation \ref{eq:limbdarkening} following \citet{pac80ugem}.

\begin{equation}
    \label{eq:limbdarkening}
    \Delta M = -2.5 \log \left[ \left( 1 + \frac{3}{2} \cos{i}\right) \cos{i}\right]
\end{equation}


\bibliographystyle{pasjtest1}
\bibliography{cvs}


\setcounter{figure}{0}
\setcounter{table}{0}
\renewcommand{\thefigure}{E\arabic{figure}}
\renewcommand{\thetable}{E\arabic{table}}

\clearpage
\section*{Supplemental information}

\begin{figure}[h]
 \begin{center}
  \includegraphics[width=\linewidth]{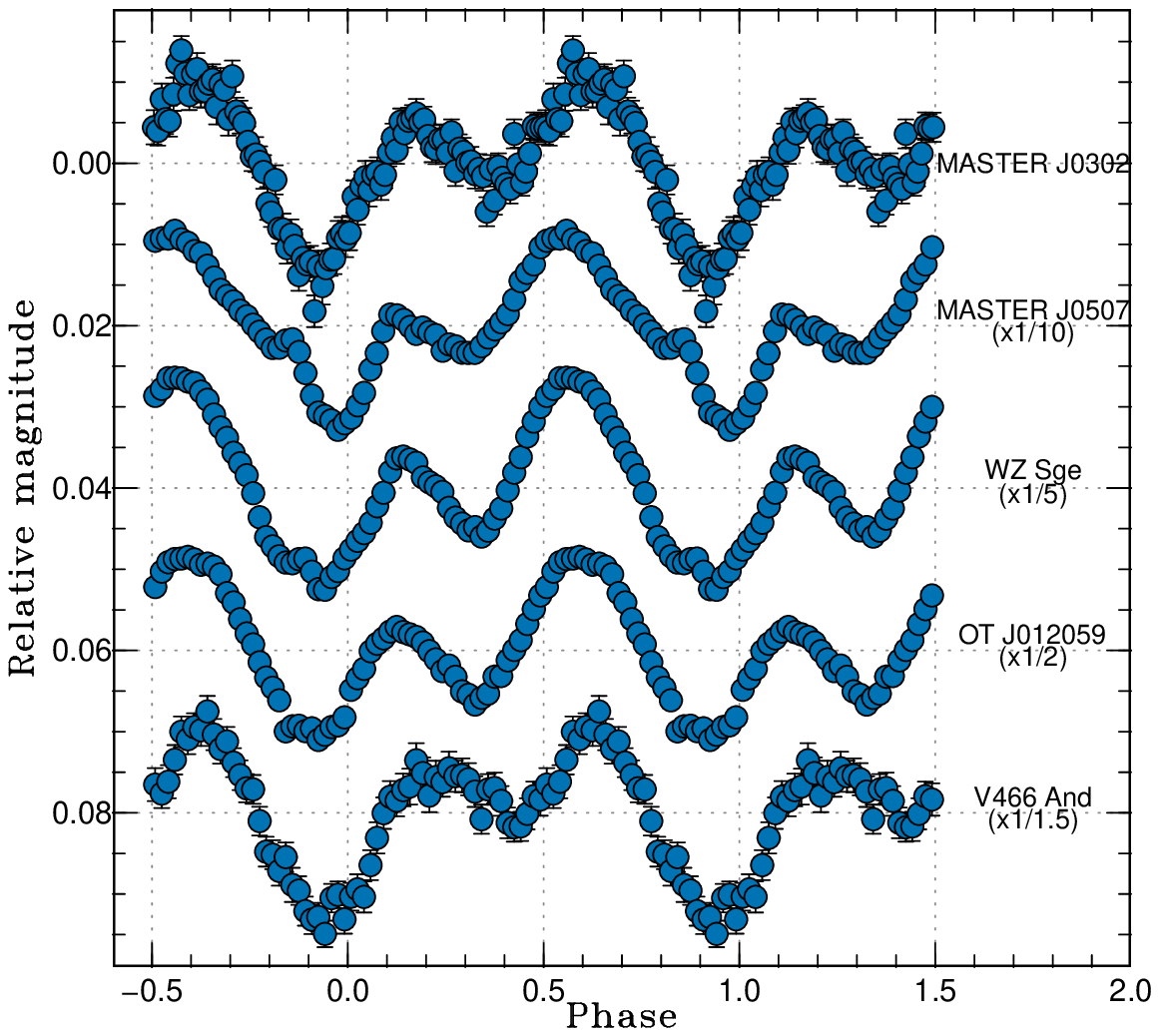}
 \end{center}
 \caption{
 The phase-averaged profile of early superhumps of MASTER OT J030227.28+191754.5 (this paper), MASTER OT J005740.99$+$443101.5, WZ Sge, OT J012059.6$+$325545, and V466 And \citep{kat15wzsge}, from upper to lower.
 The magnitude scale is multiplied and shifted for better visibility.
 }
 \label{fig:eshcomp}
\end{figure}

\begin{figure}[h]
 \begin{center}
  \includegraphics[width=\linewidth]{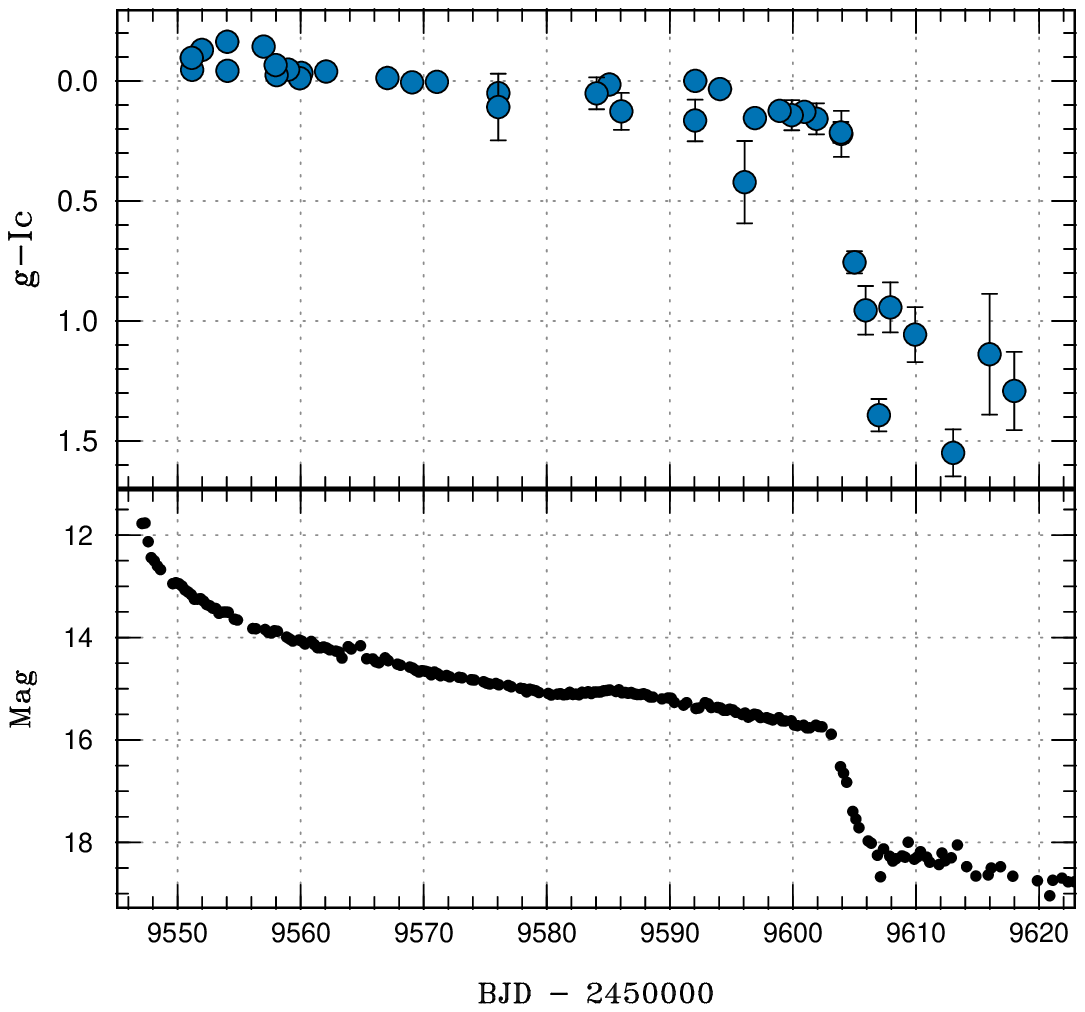}
 \end{center}
 \caption{
 The $g-I_{\rm c}$ color evolution observed with MITSuME (top) and overall $V$-band light curve in outburst (bottom).
 }
 \label{fig:giccolor}
\end{figure}

\begin{figure}[h]
 \begin{center}
  \includegraphics[width=\linewidth]{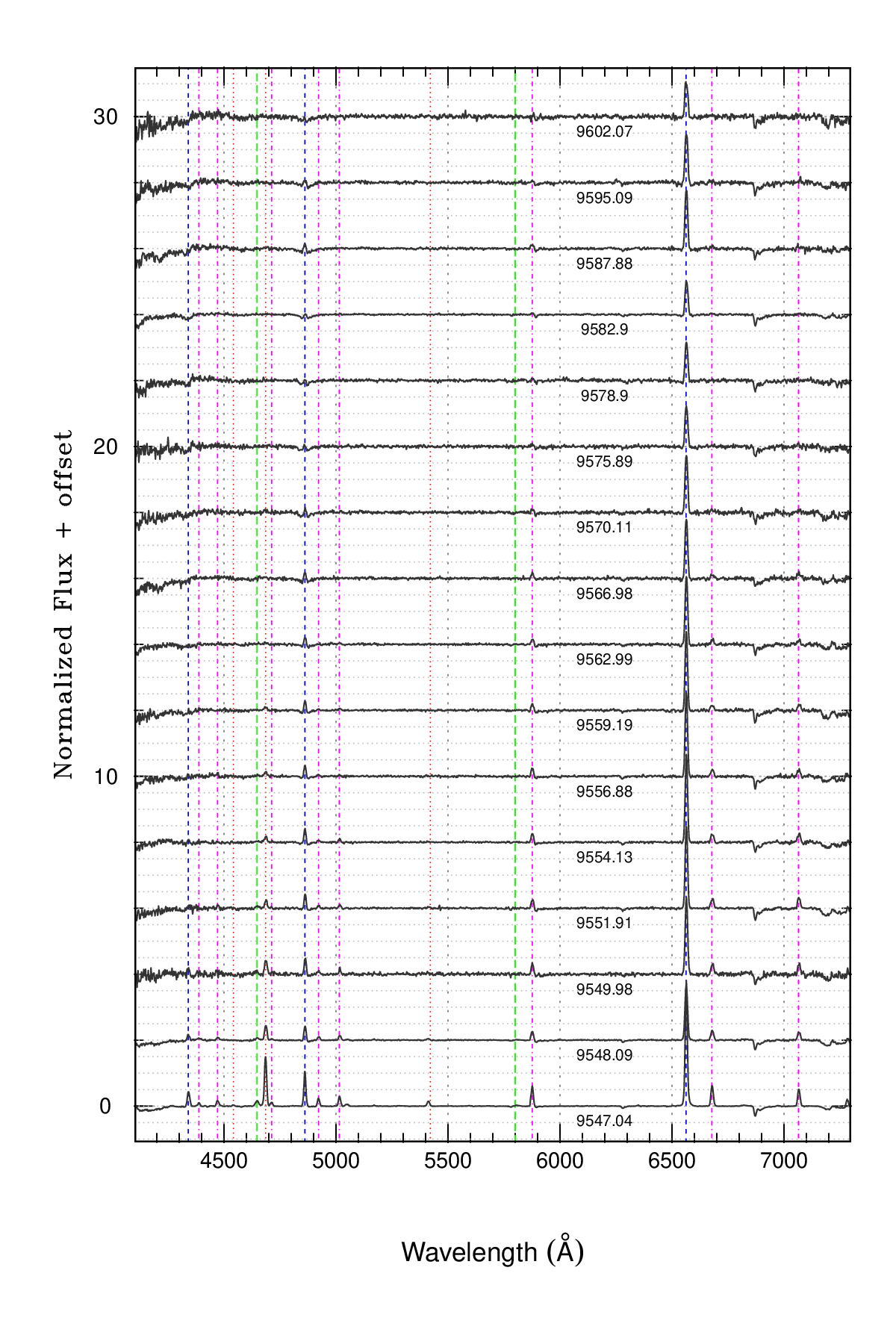}
 \end{center}
 \caption{Time evolution of normalized spectra taken with VPH-blue from lower to upper.
 The wavelength coverage is 4100--73000\AA.
 The dashed, dotted, dot-dashed, and long-dashed lines represent the wavelength of Balmer, He~\textsc{i}, He~\textsc{ii}, and Carbon/Nitrogen blends, respectively.
 The observation epochs (BJD$-$2450000) are shown below the spectra.
 }
 \label{fig:specevolline}
\end{figure}


\begin{longtable}{ccc}
  \caption{List of instruments for photometric observations}\label{tab:2}
  \hline              
    Observer code & Telescope $\&$ CCD & Observatory (or observer)\\ 
    \hline
\endfirsthead
  \hline
\endfoot
    DKS & 25cmLX200+SBIG STT-8300M & Shawn Dvorak \\
    DSO & 82cm+Apogee Alta U42 & Dark Sky Obs.\\
    FMO & 60cm+SBIG STX-16803 & Fan Mountain Obs. \\
    HHO & 1.5m Kanata telescope+HONIR &  Higashi-Hiroshima Obs. \\
    HaC & 40cmODK+Kodak 16803 & Franz-Josef HAMBSCH \\
    ISH & 105cm Murikabushi telescope+MITSuME & Ishigakijima Astronomical Obs.\\
    Ioh & 30cmSC+ST-9XE & Hiroshi Itoh\\
    KU  & 40cmSC+Apogee U6 & Kyoto U. Campus Obs.\\
    Kis & 25cmSC+QSI 632ws & Seiichiro Kiyota \\
    LCO & 100cm+Sinistro & Las Cumbres Obs. \\
    MTM & 50cmSC+MITSuME & Akeno Obs. \\
    MTO & 50cmSC+MITSuME & Okayama obs. \\
    NIC & 2m Nayuta telescope+NIC & Nishi-Harima Astronomical Obs. \\
    NKa & 60cmZeiss+FLI ML 3041 &Star\'a Lesn\'a Obs. \\
    OKU & 51cm+SBIG STXL-6303E & Osaka Kyoiku U. Obs. \\
    RFD & 35cmSC+SBIG ST8XME & Abbey Ridge Obs. \citep{rom22amateurastro} \\
    SAI & 60cm+Apogee Aspen CG42 & Sternberg Astronomical Institute \\
    SCR & 55cm SaCRA telescope+MuSaSHI & Saitama Univ. Obs. \\
    TME & 105cm Schmidt telescope+Tomo-e  & Kiso Obs. \\
    TRC & 3.8m Seimei telescope+TriCCS & Kyoto Univ. Okayama Obs. \\
    Trt & 25cm+ATIK ONE 9 & Tam\'as Tordai\\
    Van & 40cm Newton+Starlight Express SX-46 & CBA Extremadura Obs.\\
        & 40cm SC+SBIG STT-3200ME & CBA Belgium Obs.\\
    Vih & 100cm VNT Cassegrain+FLI PL1001E & Astronomical Obs. on Kolonica Saddle\\
\end{longtable}


\begin{longtable}{ccc}
\caption{Information of the comparison stars TYC2-1228.1336.1 and TYC2-1228.143.}
\label{tab:compmags}
  \hline              
       & TYC2-1228.1336.1 & TYC2-1228.143\\
    \hline
    \endfirsthead
  \hline
    \multicolumn{3}{l}{\commenta data from AAVSO website\footnote{https://app.aavso.org/vsp/}.}\\
    \multicolumn{3}{l}{\commentb data from PAN-STARRS1 \citep{panstarrs1}}\\
    \multicolumn{3}{l}{\commentc data from 2MASS \citep{2MASS}}\\
    \endfoot
    AAVSO name   & 000-BPF-626 & 000-BPF-629\\
    RA $\alpha_{\rm J2000.0}$& \timeform{03h02m28s.16342}(3) & $+$\timeform{19D19'44''.70023}(2) \\
    Dec $\delta_{\rm J2000.0}$ & \timeform{03h02m05s.52963}(1) & $+$\timeform{19D19'29''.95583}(1) \\
    $B$\commenta [mag]     & 12.08 & 14.30 \\ 
    $V$\commenta [mag]    & 11.39 & 13.24 \\ 
    $R_{c}$\commenta [mag] & 10.90 & 12.62 \\ 
    $I_{c}$\commenta [mag] & 10.45 & 12.04 \\ 
    $g$\commentb [mag]    & 11.85 & 13.66\\
    $r$\commentb [mag]    & 11.39 & 12.89\\
    $i$\commentb [mag]    & 11.20 & 12.63\\
    $z$\commentb [mag]    & 11.11 & 12.50\\
    $J$\commentc [mag]    & 10.14 & 11.38\\ 
    $H$\commentc [mag]    &  9.83 & 10.85\\ 
    $K_s$\commentc [mag]  &  9.73 & 10.74\\ 
\end{longtable}

%

\renewcommand{\thetable}{E6}
\begin{table*}
\caption{EW (\AA) of the spectral lines observed with VPH-blue.}
\centering
\label{tab:specEWs}
\begin{tabular}{ccccccccc}
  \hline              
 &  H$\alpha$ & H$\beta$ & H$\gamma$ & He~\textsc{ii} & He~\textsc{ii} & He~\textsc{ii} & C~\textsc{iii} / N~\textsc{iii} & C~\textsc{iv}/N~\textsc{iv} \\
mid BJD   &  6563 & 4861 & 4341 & 4542 & 4686 & 5411 & 4660 & 5800  \\
\hline
9547.04     & $-$46.7 (4)  & $-$11.2 (3) & $-$5.5 (2)  & $-$0.3 (1)  & $-$18.7 (2) & $-$2.0 (2) & $-$2.8 (2) & $-$0.3 (1)  \\
9548.09     & $-$23.8 (3)  & $-$4.1 (1)  & $-$1.9 (5)  & ---       & $-$6.1 (1)  & $-$0.7 (2) & $-$1.2 (2) & $-$0.2 (1)       \\
9549.99     & $-$31.5 (6)  & $-$4.6 (7)  & ---       & ---       & $-$6.0 (9)  & ---      & $-$1.5 (9) & ---       \\
9551.91     & $-$33.2 (3)  & $-$3.6 (4)  & ---       & ---       & $-$3.3 (7)  & ---      & $-$1.1 (6) & ---       \\
9554.13     & $-$34.2 (5)  & $-$3.1 (5)  & ---       & ---       & $-$2.6 (9)  & ---      & $-$0.5 (4) & ---       \\
9556.88     & $-$33.4 (3)  & $-$2.1 (3)  & ---       & ---       & $-$2.1 (9)  & ---      & ---      & ---       \\
9559.19     & $-$31.5 (5)  & $-$1.0 (6)  & 1.3 (5)   & ---       & $-$1.6 (7)  & ---      & ---      & ---       \\
9562.99     & $-$28.4 (4)  & $-$0.6 (3)  & ---       & ---       & $-$0.9 (6)  & ---      & ---      & ---       \\
9566.98     & $-$25.2 (9)  & 0.8 (4)   & ---       & ---       & ---       & ---      & ---      & ---       \\
9570.11     & $-$24.9 (5)  & 2.2 (7)   & ---       & ---       & ---       & ---      & ---      & ---       \\
9575.89     & $-$17.7 (6)  & 2.7 (5)   & ---       & ---       & ---       & ---      & ---      & ---       \\
9578.90     & $-$16.0 (5)  & 2.9 (5)   & ---       & ---       & ---       & ---      & ---      & ---       \\
9582.90     & $-$14.8 (2)  & 2.7 (4)   & 2.8 (8)   & ---       & ---       & ---      & ---      & ---       \\
9587.89     & $-$25.6 (6)  & 1.2 (4)   & ---       & ---       & ---       & ---      & ---      & ---       \\
9595.09     & $-$22.3 (1.2) & 2.0 (8)  & 2.1 (1.5) & ---       & ---       & ---      &  ---     & ---       \\
9602.07     & $-$16.4 (7)  & 4.2 (2)   & ---       & ---       & ---       & ---      & ---      & ---       \\
\hline
\end{tabular}


\begin{tabular}{cccccccccc}
 \hline
     & He~\textsc{i} & He~\textsc{i} & He~\textsc{i} & He~\textsc{i} & He~\textsc{i} & He~\textsc{i} & He~\textsc{i} & He~\textsc{i} & O~\textsc{i} \\
mid BJD & 4388 & 4471 & 4713 & 4922 & 5015 & 5876 & 6678 & 7065 & 7773 \\
 \hline
9547.04     & $-$1.1 (3)  & $-$2.3 (2)   & $-$1.5 (3)  & $-$2.6 (2) & $-$3.4 (1)   & $-$7.2 (1)  & $-$8.2 (1) & $-$6.7 (2) & $-$1.3(6) \\ 
9548.09     & $-$1.2 (9)  & $-$1.2 (4)   & $-$0.5 (1)  & $-$1.3 (3) & $-$1.7 (2)   & $-$3.5 (2)  & $-$4.4 (3) & $-$3.7 (6) & --- \\ 
9549.99     & ---       & ---        & $-$0.7 (5)  & $-$1.3 (5) & $-$1.5 (6)   & $-$4.2 (5)  & $-$4.5 (9) & $-$4.4 (7) & --- \\ 
9551.91     & ---       & ---        & ---       & $-$1.0 (6) & $-$1.1 (5)   & $-$3.6 (4)  & $-$4.4 (4) & $-$4.6 (5) & --- \\ 
9554.13     & ---       & ---        & $-$0.4 (3)  & $-$0.8 (5) & $-$0.9 (5)   & $-$3.6 (2)  & $-$3.8 (5) & $-$4.0 (7) & --- \\ 
9556.88     & ---       & ---        & ---       & $-$0.6 (4)  & ---       & $-$3.2 (3)  & $-$3.1 (5) & $-$2.6 (5) & --- \\ 
9559.19     & ---       & ---        & ---       & $-$0.4 (3)  & $-$0.8 (2)  & $-$2.6 (4)  & $-$2.4 (4) & $-$2.4 (5) & --- \\ 
9562.99     & ---       & ---        & ---       & ---       & $-$0.6 (5)  & $-$2.3 (4)  & $-$2.3 (5) & $-$1.8 (9) & --- \\ 
9566.98     & ---       & ---        & ---       & ---       & ---       & $-$2.0 (4)  & $-$1.7 (6) & $-$3.0 (8) & --- \\ 
9570.11     & ---       & ---        & ---       & ---       & ---       & $-$1.2 (6)  & $-$1.3 (8) & ---      & --- \\ 
9575.89     & ---       & ---        & ---       & ---       & ---       & ---       & ---      & ---      & --- \\ 
9578.90     & ---       & ---        & ---       & ---       & ---       & ---       & ---      & ---      & --- \\ 
9582.90     & ---       & ---        & ---       & ---       & ---       & $-$0.5 (2)  & $-$0.5 (4) & ---      & --- \\ 
9587.89     & ---       & ---        & ---       & ---       & ---       & $-$1.8 (3)  & $-$0.9 (8) & ---      & --- \\ 
9595.09     & ---       & ---        & ---       & ---       & ---       & $-$0.6 (4)  & ---      & ---      & --- \\ 
9602.07     & ---       & ---        & ---       & ---       & ---       & ---       & ---      & ---      & --- \\ 
\hline
\end{tabular}
\end{table*}

\end{document}